\documentclass[manuscript,screen]{acmart}

\AtBeginDocument{%
  \providecommand\BibTeX{{%
    \normalfont B\kern-0.5em{\scshape i\kern-0.25em b}\kern-0.8em\TeX}}}

\setcopyright{rightsretained} 

\usepackage{multirow} 
\usepackage{graphicx} 
\usepackage{color} 
\usepackage{listings} 
\usepackage{fancyvrb} 
\usepackage{newfloat} 

\lstnewenvironment{myverbatim}[1][]{%
  \lstset{
    basicstyle=\ttfamily,
    frame=tb,
    #1
  }%
}{}

\DeclareFloatingEnvironment[ fileext=lop, listname={ListofPrompts}, name=Prompt, placement=t, within=none, chapterlistsgaps=off ]{prompt}

\newcount\commentTon
\newcommand{\diego}[1]{\ifnum\commentTon=0{\color{blue} \it [DIEGO: #1]}\fi} 
\newcommand{\derek}[1]{\ifnum\commentTon=0{\color{red} \it [DEREK: #1]}\fi} 

\begin{document}

\title[Enhancing Recommendation Diversity by Re-ranking with Large Language Models]{Enhancing Recommendation Diversity by Re-ranking with Large Language Models}

\author{Diego Carraro}
\orcid{0000-0002-2857-0473}
\affiliation{
\institution{Insight Centre for Data Analytics, School of Computer Science \& IT}
\country{University College Cork, Ireland}
}
\email{diego.carraro@insight-centre.org}

\author{Derek Bridge}
\orcid{0000-0002-8720-3876}
\affiliation{
\institution{Insight Centre for Data Analytics, School of Computer Science \& IT}
\country{University College Cork, Ireland}
}
\email{derek.bridge@insight-centre.org}

\renewcommand{\shortauthors}{Diego Carraro and Derek Bridge}

\begin{abstract}
It has long been recognized that it is not enough for a Recommender System (RS) to provide recommendations based only on their relevance to users. Among many other criteria, the set of recommendations may need to be diverse. Diversity is one way of handling recommendation uncertainty and ensuring that recommendations offer users a meaningful choice. The literature reports many ways of measuring diversity and improving the diversity of a set of recommendations, most notably by re-ranking and selecting from a larger set of candidate recommendations. Driven by promising insights from the literature on how to incorporate versatile Large Language Models (LLMs) into the RS pipeline, in this paper we show how LLMs can be used for diversity re-ranking. 

We begin with an informal study that verifies that LLMs can be used for re-ranking tasks and do have some understanding of the concept of item diversity. Then, we design a more rigorous methodology where LLMs are prompted to generate a diverse ranking from a candidate ranking using various prompt templates with different re-ranking instructions in a zero-shot fashion. We conduct comprehensive experiments testing state-of-the-art LLMs from the GPT and Llama families. We compare their re-ranking capabilities with random re-ranking and various traditional re-ranking methods from the literature. We open-source the code of our experiments for reproducibility. Our findings suggest that the trade-offs (in terms of performance and costs, among others) of LLM-based re-rankers are superior to those of random re-rankers but, as yet, inferior to the ones of traditional re-rankers. However, the LLM approach is promising. LLMs exhibit improved performance on many natural language processing and recommendation tasks and lower inference costs. Given these trends, we can expect LLM-based re-ranking to become more competitive soon.
\end{abstract}

\ccsdesc[500]{Information systems~Information retrieval diversity}
\ccsdesc[500]{Information systems~Language models}
\ccsdesc[500]{Information systems~Recommender systems}

\keywords{Recommender Systems, Large Language Models, Diversity, Re-ranking}


\maketitle

\section{Introduction}\label{sec:introduction}

Large Language Models (LLMs) have rapidly become a breakthrough technology since the advent of their popular pioneers GPT \cite{radford2018GPT} and BERT \cite{devlin2018bert}, introduced in 2018. Since then, many more LLMs with enhanced capabilities have been proposed, such as ChatGPT, Bard and Llama. They can perform various language-related tasks, including, for example, translation, summarization and conversation; and they give the appearance of understanding complex contexts and exhibiting reasoning, planning and problem-solving capabilities \cite{kojima2022reasoners, valmeekam2023planning, yao2024tree_solving}. LLMs are typically pre-trained with large datasets of text to serve as general-purpose models and then adapted to different downstream tasks and domains by a variety of approaches \cite{min2023llms} --- prompting, Supervised Fine Tuning (SFT) and  Reinforcement Learning from Human Feedback (RLHF) being among the most widely used. Prompting refers to giving the LLM a specific input (e.g.\ a question, a sentence, a set of instructions), known as a prompt, designed to drive the LLM to generate a desired output. Supervised Fine-Tuning adapts the model by further training the LLM on a task-specific labelled dataset. Reinforcement Learning from Human Feedback (RLHF) \cite{ziegler2019rlhf} fine-tunes the model using human preference data and is typically used in settings where large amounts of labelled training data are not available or where the domain requirements are human-centred.

Given their abilities and their versatility, researchers in Recommender Systems (RSs) have proposed different ways of integrating LLMs into a recommendation pipeline, for example, to enhance data augmentation, perform feature engineering, design scoring functions, directly provide recommendations, and furnish explanations of recommendations \cite{lin2023survey_llms_as_recs, fan2023recommender_llms}. Some researchers have taken a unified approach, where the entire recommendation task is framed as a text-generation task: a single LLM can perform all the functionalities mentioned above (e.g.\ \cite{geng2022P5, cui2022m6}), showing how well LLMs can generalise across different recommendation settings. 

All the work on RSs we have cited focuses on providing recommendations that are as relevant as possible to a user. However, long before the use of LLMs in RSs, the RS literature recognized that relevant recommendations might not be enough to satisfy a user fully: providing a set of recommendations that is diverse (different from one another) or recommendations that are novel (new or previously unexplored), serendipitous (surprising, unexpected, and going beyond the user's typical preferences) and fair (not discriminating between certain groups of users or items) is also crucial, see, e.g., \cite{vargas2015thesis, kaminskas2016diversity, wang2023survey_fairness}. Strategies that optimize for these so-called beyond-accuracy objectives have been proposed for traditional RSs, while the topic is under-investigated for LLM-based RSs. The goal of this paper is to contribute to filling this gap. 

Our work investigates enhancing the top-$n$ recommendation diversity by re-ranking using general-purpose state-of-the-art LLMs. Other works 
propose to re-rank recommendations leveraging LLMs to improve relevance, but to the best of our knowledge, we are the first ones focusing on beyond-relevance objectives. In particular, our contributions are tailored to the following research questions:

\begin{itemize}
    \item \textit{RQ1: Can LLMs interpret the building blocks of RS diversification re-ranking, i.e.\ item diversity and item re-ranking?} 
    \item \textit{RQ2: Given an initial candidate set of items ranked by their relevance to a user, can general-purpose LLMs re-rank these items to increase their recommendation diversity?}
    \item \textit{RQ3: How do different zero-shot prompts affect the re-ranking performance in terms of relevance and diversity?}
    \item \textit{RQ4: What are the trade-offs of LLM-based and traditional greedy diversification re-rankers in terms of performance, costs and other important aspects such as control over data, memory footprint and generalisation across domains, among others?}
\end{itemize}

To answer RQ1, we run an informal preliminary study where, first, we ask ChatGPT to perform some re-ranking tasks, supplementing existing positive evidence already provided in the information retrieval \cite{sun2023chatgptgoodatsearch} and RS \cite{hou2024llm_rankers} domains. Then, we prompt the model to compare two lists of items in terms of their item diversity. Encouraged by positive findings in these informal studies, where the LLM exhibits re-ranking capabilities and knowledge of item diversity, we then propose a more rigorous methodology to answer RQ2-4. 

First, we frame diversification re-ranking as a text generation task in which we prompt an LLM in a zero-shot fashion to generate a more diverse ranking from a candidate ranking. We design eight prompt templates that contain different instructions for re-ranking to explore different ways to approach the generation. We explore a wide range of templates that differ in their complexity, from simple high-level task descriptions to more advanced ones where we provide additional information about the items and further re-ranking guidelines. We also introduce ways of addressing LLM hallucinations (i.e.\ where the model produces invalid recommendations), a common problem that is usually ignored by most of the LLM literature on RSs. 

Then, we run experiments on an anime movie dataset and on a book dataset where a Matrix Factorization RS optimized for relevance provides candidate recommendations. We apply our proposed LLM-based diversification re-ranking to these recommendations using state-of-the-art LLMs. Specifically, we use proprietary models from OpenAI\footnote{\url{https://openai.com/}} (namely, ChatGPT and InstructGPT) and also open-source models from Meta's Llama family\footnote{\url{https://llama.meta.com/}} (namely, Llama2-7B-Chat and Llama2-13B-Chat). We assess to what extent recommendation diversity improves and how relevance is affected (RQ2, RQ3). We compare with a random re-ranker and some traditional state-of-the-art greedy strategies that intelligently re-arrange items to balance relevance and diversity.

To address RQ4, we run a comprehensive analysis beyond performance for both approaches, where we discuss their trade-offs in terms of costs, control over data, and memory footprint, among other aspects.

Our results demonstrate that, in both datasets, the LLM-based re-rankers we proposed can interpret the re-ranking task, sacrificing some relevance to improving diversity. While LLM-based re-rankers achieve better relevance/diversity trade-offs than random re-ranking, they are still inferior to traditional re-rankers, particularly in relevance-aware metrics, due to issues such as drawing lower-ranked items and the presence of random recommendations. OpenAI's models, particularly ChatGPT, outperform Meta's models, such as Llama2-13B-Chat, aligning with existing literature that credits OpenAI models with more advanced capabilities. The effectiveness of different prompt templates varies, indicating the need for tailored prompt design, with feature-aware templates generally performing better. Finally, the traditional greedy approach are faster and less resource-intensive, making the costs and constraints of LLM-based re-ranking a significant barrier for practical use. Nevertheless, our study shows that LLM-based has considerable promise. It will become more competitive as LLMs improve and as the costs of using them continues to fall.

The remainder of this paper is structured as follows. Section \ref{sec:related_work} reviews related work; Section \ref{sec:diversification} presents both the preliminary study and our LLM-based re-ranking approach; Section \ref{sec:experiments} describes experiments and Section \ref{sec:results} analyses their results. Section \ref{sec:beyond_performance} offers a broader perspective on the trade-offs of LLM-based and greedy re-ranking approaches. Finally, \ref{sec:conclusions} wraps up our findings, discusses the limitations of our study and outlines future directions.

\section{Related work} \label{sec:related_work}

\subsection{Measuring recommendation diversity}
The most common way to measure the diversity of a set of items, such as a set of $n$ recommendations, uses item dissimilarity. In particular, the average \cite{smyth2001similarity} or aggregate \cite{ziegler2005improving} pairwise distance between items in the set are popular diversity measures of this kind. 
Different ways of quantifying item distance have been used. In content-based systems, for example,  the complement of the Jaccard similarity on sets of features such as genres \cite{vargas2011rank}, the complement of the cosine similarity on term vectors \cite{ekstrand2014user}, and distance in a taxonomy \cite{ziegler2005improving} are possible methods. When items have vectors of user ratings, item distance can be computed using the complement of Pearson correlation \cite{vargas2011rank}, the complement of cosine similarity \cite{ribeiro2012pareto} or Hamming distance \cite{kelly2006enhancing}. Item distance can also be computed on the latent item feature vectors in Matrix Factorization approaches \cite{vargas2011intent, shi2012portfolio, su2013set}.

Some researchers have argued that the traditional diversity metrics do not correlate well with users' perceptions and have proposed alternatives, especially when item features are based on item genres. For example, these new metrics also account for coverage of item feature values, redundancy and the size of the recommendation list \cite{vargas2014coverage}. 

There are also metrics that consider both diversity and relevance at the same time: they quantify the trade-off between diversity and relevance with a single metric. $\alpha$-NDCG, for example, was introduced by \cite{clarke2008novelty} to evaluate information retrieval systems, where the score of a retrieved document is penalized if it shares features with documents ranked higher in the list of retrieved documents. This metric has also been used in RS research in \cite{carraro2020active, kaya2019subprofile}. Inspired by $\alpha$-NDCG, Vargas \& Castells weight a diversity score with item relevance, under the assumption that diversity on irrelevant items does not play a major role in satisfying a user \cite{vargas2011rank}. They designed a framework that allows one to define several metrics that account for a diversity and relevance trade-off based on a ranking discount. In our experiments, we use genres as item features, and we employ their framework and some metrics from it, such as $\alpha$-NDCG, Expected Intra-List Diversity, and relevancy-aware Subtopic Recall; see Section \ref{sec:eval_metrics}.

Finally, for completeness, we note that metrics have also been defined for other types of diversity, such as temporal diversity metrics that capture the extent to which recommendations generated for the same user differ over time, or metrics that take a system-centric, rather than user-centric, perspective by analysing how recommendations differ across users. We refer to \cite{kaminskas2016diversity} and \cite{castells2021novelty} for comprehensive surveys of diversity-aware metrics.

\subsection{Increasing recommendation diversity}\label{sec:increasing_div}

As well as \emph{measuring} diversity, researchers have proposed ways of \emph{increasing} the diversity of a set of recommendations. Diversification methods can be categorised into two main groups: diversity modelling and diversity post-processing. 

The first group of works includes recommendation models that are directly optimized to produce diverse recommendations. Most of these RSs are derivations or extensions of the popular and effective Matrix Factorization (MF) model. For example, Shi et al.\ \cite{shi2012portfolio} designed an objective function that balances the expected relevance and variance of the recommendation list. The variance level is obtained from the latent factor vectors of an MF model and correlates with a user's spectrum of tastes. In other words, the latent factors of a user who rates
diverse items have higher variance compared to a user who rates similar items; the former user will consequently receive more diversified recommendations from the model than the latter user does. Another example is \cite{hurley2013personalised}, which modified the learning function of a pairwise learning-to-rank approach (for implicit feedback datasets) to account for item dissimilarities. A third example is \cite{su2013set}, which proposed a revised pairwise learning-to-rank model that works at the item set level (rather than the individual item level): their model is trained by comparing item sets using both relevance and diversity criteria to establish the pairwise ranking.

The second group of works includes approaches that apply a post-processing step (typically, using a greedy strategy) that refines an initial list of recommendations from a conventional RS to obtain a final set of recommendations that balances relevance and diversity. More formally, these techniques typically produce a final list of recommended items $RL$ of size $n$ from a larger set of candidate recommendations $CL$ of size $m$ (with $m > n$), where $CL$ is generated by a baseline recommendation algorithm that ranks items based on their relevance. The vast majority of the RS literature on this uses greedy re-ranking, which involves an iterative process where items are selected from $CL$ one by one and added to $RL$. At each step, the greedy algorithm selects the candidate item that maximizes a re-ranking objective function $f_{obj}$, like the one defined in Eq.\ \ref{formula:greedy_reranking}. Here, $f_{obj}$ is a linear combination of two scores, i.e.\ a candidate item’s relevance and the diversity this item brings when added to the current version of $RL$. In particular, in the equation, $rel(i)$ denotes the relevance score of the item $i$ to a user. $div(i, RL)$ denotes the diversity of $RL$ when the item $i$ is included in $RL$ and it typically makes use of a distance function between items. Finally, the hyperparameter $\lambda \in [0,1]$ controls the trade-off between the influence of relevance and diversity during the re-ranking.

\begin{equation}\label{formula:greedy_reranking}
    f_{obj}(i, RL) = \lambda \cdot rel(i) + (1 - \lambda) \cdot div(i, RL)
\end{equation}

Greedy re-ranking has been widely adopted in the RS literature because it is easy to implement and integrate into existing recommender system pipelines. Indeed, its relevance $rel(i)$ and diversification $div(i, RL)$ components can be adapted based on the application domain, and their trade-off can be explicitly controlled with $\lambda$. 

Maximal Marginal Relevance (MMR) \cite{carbonell1998MMR} is one of the first diversification techniques of this kind. It was originally proposed for the information retrieval domain but later adopted in the RS literature, e.g.\ \cite{vargas2011intent, dinoia2017adaptive}. In MMR, $div(i, RL)$ is computed as the negative of $i$'s maximum similarity to
items already in $RL$.

In a case-based recommender scenario, Smyth \& McClave implemented $rel(i)$ as the similarity between the user’s query and an item or case $i$, while $div(i, RL)$ was the average pairwise distance between $i$ and the items or cases already in $RL$, the distance being the complement of their similarity \cite{smyth2001similarity}. In a conversational recommender scenario, Kelly \& Bridge took $rel(i)$ as the predicted item’s relevance and $div(i, RL)$ is again the average pairwise distance between $i$ and the items in $RL$, but this time using the normalized Hamming distance of the two items’ binary rating vectors \cite{kelly2006enhancing}.

A more recent stream of research arose from Vargas' work \cite{vargas2015thesis}, which proposed greedy re-ranking strategies inspired by Intent-Aware (IA) diversification from the field of information retrieval \cite{agrawal2009diversifying}. In essence, IA methods in information retrieval seek to optimize the relevance and diversity trade-off by retrieving documents that are relevant to the user's query but which also cover different aspects of the query (e.g.\ different meanings of an ambiguous query). In the RS domain, IA methods seek to retrieve items that are predicted to be relevant to the user while covering different aspects such as item feature values or user tastes. For example, Vargas presented explicit Query Aspect Diversification (xQuAD) \cite{vargas2011intent} and its variant Relevance-based xQuAD (RxQuAD) \cite{vargas2012explicit}. xQuAD focuses on maximizing the probability that a user will \textit{select} at least one of the recommended items by building an $RL$ that covers all the different user aspects. RxQuAD introduces a relevance model into xQuAD to maximise relevance rather than the probability of choosing a single item. 

There are also works that perform post-processing but do not use greedy re-ranking. Instead, they typically aim at solving optimization problems to find the optimal ranking for balancing relevance and diversity; some examples are \cite{zhang2008avoiding, jambor2010optimizing, ribeiro2012pareto}. 

In the experiment described in this paper, we use  MMR, xQuAD and RxQuAD as our greedy re-ranking baselines against which we compare the performance of our LLM-based re-rankers (see Section \ref{sec:experiments}).

\subsection{LLMs and recommender systems}


Broadly speaking, work on using LLMs in the RS domain can be categorized into \textit{supporting recommendations} or \textit{directly providing recommendations}. 

Representative examples from the first group of works use an LLM to augment training data, extract features, and explain recommendations. Borisov et al.\ \cite{borisov2022language_tabular} exploit LLMs to generate synthetic examples from tabular data; Liu et al.\ \cite{liu2023first_new} prompt an LLM to craft news summaries, user profiles, and synthetic news for a news recommender; Mysore et al.\ \cite{mysore2023augmented_narrative} propose an LLM-based pipeline that converts user-item interaction data, commonly used to train collaborative filtering systems, into datasets to train narrative-driven recommenders. Inspired by the Retrieval-Augmented Generation (RAG) paradigm \cite{gao2024rag}, Di Palma \cite{dipalma2023rag_rec} advocates a comprehensive approach, which he calls a Retrieval-augmented Recommender System, which combines the strengths of retrieved information about items with LLMs to enhance the recommendation task.

When used as a feature extractor, an LLM is typically fed with textual user features, item features or user-item interaction data (e.g.\ user behaviour or reviews) and returns vectorized embeddings that can be used by a traditional recommender model (often a neural network) to perform various recommendation tasks. Features are extracted, for example, from code snippets \cite{rahmani2023code_rec}, from news \cite{zhang2021unbert_news}, from tweets \cite{zhang2022twhin_tweets}, and from documents \cite{liu2021baidu_search}, and applied to enhance cross-domain recommendation \cite{ding2021zero_cross, hou2022towards, hou2023learning_cross}.

As they can generate fluent and coherent text, LLMs are also powerful tools used to provide explanations in natural language for different kinds of recommendations, e.g.\ for products in e-commerce \cite{chen2021generate_explanations, cui2022m6}, for leisure products \cite{li2023personalized_explanations}, and for music, video games and food \cite{hada2021plug_play_explanation}.

Works in the second group use LLMs to provide recommendations directly, either by generating item suggestions or by giving scores to user-item pairs. When LLMs generate item suggestions, they can be in the form of item names \cite{li2023gpt4rec, cui2022m6} or identifiers \cite{geng2022P5, petrov2023generative}; and the LLM can make next-item recommendations \cite{wang2023zero, geng2022P5} or it can rank a set of candidate items \cite{cui2022m6}. When LLMs generate scores, they can be in the form of predicted ratings \cite{geng2022P5,li2023sequential}, click-through rates and conversion rates \cite{cui2022m6}, or preferences \cite{zhang2023prompt, tallrec}.

Other works, different from the two groups we have reviewed so far, are more related to re-ranking recommendations, akin to what we propose in this paper. We briefly review them in the following and we draw explicit comparison with our work in Section \ref{sec:diversity_reranking}. Dai et al.\ \cite{dai2023uncovering}, Li et al.\ \cite{li2023preliminary_news} and Hou et al.\ \cite{hou2024llm_rankers} all use several models from the ChatGPT family to rank an unordered candidate set of items in a zero-shot scenario for domains such as movies, books, music and news, additionally leveraging historical user interactions in the prompts. Sun et al.\  \cite{sun2023chatgptgoodatsearch} investigate the capabilities of ChatGPT and GPT-4 at re-ranking candidate documents retrieved by a traditional information retrieval system (in a zero-shot generation and by distilling a specialized model).  Suzgun et al.\ \cite{suzgun2022prompt_rerank} use LLMs for textual style transfer: the system first generates a set of candidate texts in the target style (by zero-shot or few-shot prompting with GPT2) and then re-ranks them according to three types of scores provided by different LLMs.
All these works show promising results for LLM-based re-ranking (especially for ChatGPT) but they all struggle to outperform state-of-the-art competitors. Moreover, there is no consistent methodology for evaluating LLM-based re-rankers (taking into account, e.g., invalid recommendations). In Section \ref{sec:diversity_reranking}, we will compare in detail our work with this literature, highlighting similarities and differences for each component of our system.


\section{Recommendation diversification using LLMs for re-ranking}
\label{sec:diversification}

Our work investigates recommendation diversification by using an LLM to re-rank a set of pre-computed candidate recommendations. We frame this task as a zero-shot text generation task and discuss more advanced approaches in Section \ref{sec:conclusions}. For re-ranking, we use and compare two state-of-the-art families of LLMs, i.e.\ ChatGPT/InstructGPT and Llama-2-Chat, which, using SFT and RLHF, have been fine-tuned from their foundation models (GPT and Llama respectively) to follow instructions. 

\subsection{A preliminary study}\label{sec:preliminary_study}

Although there is extensive evidence that both ChatGPT and Llama-2-Chat can successfully perform many different tasks, to the best of our knowledge, little is known about whether they are good at re-ranking and whether this can be done in the context of diversity in RSs. The closest works to ours that offer some positive evidence in these regards are \cite{li2023preliminary_news, dai2023uncovering}, which explore ranking an unordered candidate set of items; and \cite{sun2023chatgptgoodatsearch}, where the task is re-ranking a candidate list of items previously ranked by a retrieval system. These works focus on ChatGPT only. To the best of our knowledge, there is no published research to assess whether LLMs are capable of reasoning about item diversity. Thus, we begin our informal investigations by answering RQ1. More specifically, we find answers to the following:

\textit{Q1 Can LLMs interpret and perform re-ranking tasks?}

\textit{Q2 Do LLMs have knowledge about item diversity?}

We note that this preliminary study is rather informal and is not a comprehensive evaluation of an LLM's re-ranking capabilities. Instead, it serves as an indication of their potential capabilities towards our larger goal. 

To answer Q1, we designed three simple prompts asking the LLMs to re-rank a list of the titles of 20 anime movies. (We use anime movies for this preliminary study because they come from one of the datasets we use in the more substantial experiments described in Section \ref{sec:experiments}.) Specifically, we ask the LLMs to re-rank the lists so that they are ordered alphabetically, or by popularity, or by release date. We expect that the complexity of the tasks should affect the re-ranking performance: when re-ranking by popularity and release date, an LLM must leverage additional details about the items to accomplish the task, whereas alphabetical re-ordering can be done by using the items' titles only.

In Appendix \ref{sec:appendix_prelim}, we show three representative examples of the prompts and ChatGPT's outputs. (Llama-2-Chat's outputs are similar, so we do not show them.) As these examples demonstrate, ChatGPT can perform the re-ranking tasks with different degrees of accuracy. In particular, alphabetical re-ranking (Q1-P1) is the most accurate, with only one item incorrectly ranked. By contrast, popularity (Q2-P2) and release date (Q3-P3) re-ranking are only partially consistent with the ground-truth rankings. We ran several examples of this kind (with different item lists to re-rank). We compared the ChatGPT outputs with ground-truth rankings. We obtained scores that vary between 0.5 and 0.8 for popularity re-ranking and between 0.35 and 0.7 for release date re-ranking, according to the rank-biased-overlap metric \cite{webber2010rbo}. 

To answer Q2, we designed one prompt that asks the LLM to indicate which of two lists, each composed of 10 anime titles, is the more diverse. We deliberately do not provide any definition of diversity in the prompt so that the LLM must interpret its meaning for itself. Similarly to Q1, we built examples that exhibit different degrees of complexity: an ``easier'' prompt that compares list A that covers one genre (e.g.\ all items are comedies) and list B that covers multiple genres; and a ``harder'' prompt that compares two lists that both cover multiple genres. Again, we ran several examples of this kind (with different item lists to re-rank).

In Appendix \ref{sec:appendix_prelim}, we show two representative examples of the prompts and ChatGPT's outputs. Q2-P1 is the easier prompt, while Q2-P2 and Q2-P3 are the harder prompts. (Again, Llama-2-Chat's outputs are similar, so we do not show them.) In each example, ChatGPT's output first includes its interpretation of the meaning of diversity (i.e.\ a genre-based definition), followed by the candidate lists augmented with the genres it associates with each item in the list, concluding with the answer to the prompt's question along with an associated explanation. 

As expected, for Q2-P1, ChatGPT correctly indicates that list B is the more diverse, and it justifies its answer from a genre-based diversity perspective, i.e.\ the number of different genres a list covers.

For Q2-P2, ChatGPT indicates that list A is the more diverse and again justifies its answer from a genre-based diversity coverage perspective. According to the genres that ChatGPT assigns to the lists, its answer is correct (list A covers 15 genres, and list B 13 genres). However, if we extract ground-truth genres from the dataset that we use in Section \ref{sec:experiments}, then list B is more diverse (list A covers 18 ground-truth genres, and list B 19). 

Finally, For Q2-P3,  ChatGPT indicates that list B is the more diverse (for the same reason above). However, ChatGPT's conclusion does not correspond with the genres that it assigns because, in fact, it assigns 11 genres to list B against 12 for list A. 
The incorrectness of the answer is also confirmed by the ground-truth genres we extracted from the dataset, i.e.\ 19 for B and 24 for A.

This preliminary study suggests that an LLM can re-rank lists of items and can provide insights into their diversity, using its embedded knowledge (without any additional fine-tuning), albeit imperfectly. Moreover, the LLM showed some degree of reasoning when performing the tasks and was able to provide highly structured answers,. Correctness also seems correlated with task difficulty. 

Again, we emphasis that this preliminary study is not rigorous, and its findings are not measures of LLM abilities. In particular, we loosely defined ``correctness'' when evaluating answers to Q2. 
What we take from this informal investigation is an indication that it is worth proceeding to the more rigorous study that we present in the remainder of this paper.

\subsection{Prompt-based zero-shot diversification re-ranking}\label{sec:diversity_reranking}

Following the previous section's largely positive findings, we propose a methodology to answer the research question RQ2 presented in Section \ref{sec:introduction}, i.e.\ whether LLMs can enhance recommendation diversity by re-ranking. We consider the greedy re-ranking framework presented in Section \ref{sec:increasing_div}, where the main components are the candidate list $CL$, the final recommendation list $RL$, a relevance function $rel$ defined over the items in $CL$, a diversification function $div$ defined over the items in $CL$ and the list $RL$, and a hyperparameter $\lambda$  to control the relevance/diversification trade-off. We also assume the candidate list $CL$ is produced by a baseline recommender. In Section \ref{sec:experiments}, we choose a Matrix Factorization model to provide $CL$ for our experiments, but in practice this can be done by any recommender that can produce a ranked list of items.

We frame the re-ranking task as a text generation task to be accomplished by an LLM in response to specific prompts that request the production of $RL$ from $CL$. The prompt template comprises three sections, see Prompt \ref{prompt}. The first section of the prompt provides the context and instructions the LLM must follow to generate $RL$. The second section provides the output format the LLM must adhere to. The third section provides the candidate list $CL$ that the LLM must re-rank. However, except for the use of triple backticks to delimit $CL$, we do not explicitly mark-up the three sections: the tags in Prompt \ref{prompt} are for the purposes of exposition in this paper, they are not included in the final prompt. The decision to exclude them is in line with what the literature suggests (e.g.\ \cite{dai2023uncovering, sun2023chatgptgoodatsearch, geng2022P5}). We will now discuss each of the three sections in more detail.

\begin{prompt}[t]
\begin{myverbatim}[]
<Instructions>:
You are given a ranked recommendation list of {m} items for a user, 
delimited by triple backticks.
Your task is to re-rank this candidate list and provide a final top-{n} 
recommendation list where the goal is to {goal string}. Strictly use the 
following format for the output, and don't provide additional information.

<Output format>:
1-> <item name>
2-> <item name>
...
{n}-> <item name>

<CL>:
```
1. {candidate item name ranked #1 in candidate set} {additional_features}
2. {candidate item name ranked #2 in candidate set} {additional_features}
...
{m}. {candidate item name ranked #m in candidate set} {additional_features}
```
\end{myverbatim}
\caption{Generic re-ranking prompt template. Tags in angular brackets delimit sections (but they do not appear in the final prompt), and placeholders for data are enclosed in braces.} \label{prompt}
\end{prompt}

\begin{table}[t]
\caption{Mapping of placeholders in the generic prompt template.}
\begin{tabular}{|c|c|c|c|}
\hline
\multicolumn{1}{|l|}{\textbf{Template id}} & \multicolumn{1}{c|}{\pmb{$\lambda$}} & \textbf{\{goal\_string\}}                                                                                                                                                                                                & \textbf{\{additional\_features\}}            \\ \hline
T1                                & 0.5                      & ``balance relevance and diversity''                                                                                                                                                                               & None                                \\ \hline
T2                                & 0                        & ``maximize the items' diversity in the list''                                                                                                                                                                     & None                                \\ \hline
T3                                & 0                        & ``maximize the items' genre-based diversity in the list''                                                                                                                                                         & None                                \\ \hline
T4                                & 0.5                      & ``balance relevance and genre-based diversity''                                                                                                                                                                   & None                                \\ \hline
T5                                & 0.5                      & ``balance relevance and diversity''                                                                                                                                                                               & List of the candidate item's genres \\ \hline
T6                                & 0.5                      & ``balance relevance and genre-based diversity''                                                                                                                                                                   & List of the candidate item's genres \\ \hline
T7                                & 0.5                      & \begin{tabular}[c]{@{}c@{}}``balance relevance and diversity. \\ Guidelines to perform the re-ranking are: \\ Use the plot summary information of each \\ item attached in curly bracke''\end{tabular}          & Candidate item's plot summary             \\ \hline
T8                                & 0                        & \begin{tabular}[c]{@{}c@{}}``maximize the books' diversity in the list.\\ Guidelines to perform the re-ranking are: \\ Use the plot summary information of each \\ item attached in curly bracket''\end{tabular} & Candidate item's plot summary             \\ \hline
\end{tabular}
\label{table:mapping_templates}
\end{table}

Clear instructions are crucial to obtaining the desired output since the LLMs we use are aligned with high-quality instruction-based prompts. We designed eight subtypes of the generic template, each identified by a different re-ranking goal; see Table \ref{table:mapping_templates}. Each goal maps to a different $\lambda$ value, with T1, T4-T7 aiming at equally balancing relevance and diversity (i.e.\ $\lambda = 0.5$ in traditional greedy re-ranking approaches) and T2, T3, T8 aiming at maximizing diversity only (i.e.\ $\lambda = 0$). The related research literature proposes alternative ways to model different re-ranking trade-offs. Li et al.\ \cite{li2023preliminary_news} address the trade-off between recommendation relevance and fairness by explicitly instructing the LLM to select a specific number of candidate items from different groups of items --- in their case, popular and unpopular items. Suzgun et al.\  \cite{suzgun2022prompt_rerank} use a combination of scores from different LLMs to adjust the trade-off between textual similarity, target style strength, and fluency for arbitrary textual style transfer.

The prompt template also includes placeholders \{m\} and \{n\}. These are replaced with the size of $CL$ and $RL$. We explain how we choose the values of  $m$ and $n$ in Section \ref{sec:experiments}.

The prompt template specifies an output format to guide the LLM's generation, simplify output parsing (i.e.\ extracting the final ranking from the LLM's answer), and reduce its ambiguities. Similarly to \cite{sun2023chatgptgoodatsearch}, we opted for a numbered list (with numbers representing an item's rank, i.e.\ its relevance in $CL$), followed by the string \verb|->|, to help to identify item names when parsing the output. To the best of our knowledge, the literature offers no advice about how to represent ranked lists in a prompt. There have, however, been works in the literature that represent un-ranked candidate lists in prompts, e.g.\ by identifying items with letters \cite{dai2023uncovering} or by simply appending items one after another \cite{li2023preliminary_news}.

Finally, the prompt must include the list $CL$. We represent its items with their names, e.g.\ anime titles or book titles (see Section \ref{sec:experiments}), similar to what is done in most of the literature (although some works use item identifiers, e.g.\ \cite{geng2022P5}). Items are also numbered in our prompts to report their rank (i.e.\ their relevance). Also, for templates T5, T6, T7 and T8, which we call feature-aware templates, we replace the \verb|{additional_features}| placeholders to augment an item with its genres (T5, T6) and its description (T7, T8), attempting to equip the LLM with explicit item features to leverage for re-ranking. (In our experiments in Section \ref{sec:experiments}, we use genres extracted from the datasets and descriptions extracted from prompting ChatGPT.) Feature-aware templates can be seen as a form of RAG-based prompting, where each prompt is augmented with some extra contextual information. The only difference between traditional RAG and our approach is that traditional RAG systems typically retrieve this additional context on-the-fly (i.e.\ when generation occurs), while we retrieve the whole set of item features in advance of re-ranking (see Section \ref{sec:experiments}). In Appendix \ref{sec:appendix_rerank}, we show an example of a prompt (i.e.\ after the placeholders in the template have been instantiated). We also show ChatGPT's output. 

Templates T1-T8 explore different degrees of complexity and help us in addressing RQ3. However, several other prompting techniques might have been explored --- for example, few-shot prompting (i.e.\ providing task-related demonstrative examples), or Chain-Of-Thoughts (CoT) prompting (i.e.\ where the LLM is driven to provide the requested output through intermediate steps) \cite{liu2023pre_prompt, wei2022cot}. However, the main goal of this study is to shed light on the capabilities of LLMs to understand the diversification re-ranking task and produce meaningful recommendations rather than finding the prompting strategy that performs best. The exploration of further more advanced LLM-based re-ranking strategies is left as future work.

Finally, in our prompt-based formulation, we do not explicitly model $f_{obj}$, nor the $rel$ or $div$ functions, in the prompts. We instead `implement' $f_{obj}$ by including key terms such as ``relevance'' and ``diversity'' (or ``genre-based diversity'') for the $rel$ and $div$ functions, respectively, and terms such as ``balance'' and ``maximize'' to model the $\lambda$ trade-off. However, we posit that our prompt-based implementation is a raw form of greedy re-ranking. This is likely because LLMs such as ChatGPT and Llama2 are autoregressive models that generate text (i.e.\ the final list $RL$) in an iterative way, i.e.\ item by item until $RL$ is composed of $n$ elements.

\section{Experiments}\label{sec:experiments}

In this section, we outline the experimental methodology and the datasets we used in our experiments. We have open-sourced the code for reproducibility.\footnote{\url{https://github.com/cdiego89phd/chat-reranking}} 

\subsection{Datasets and preprocessing}
In our experiments, we use the Anime Recommendation Database\footnote{\url{https://github.com/Hernan4444/MyAnimeList-Database}} and the Goodreads Book Graph Datasets\footnote{\url{https://mengtingwan.github.io/data/goodreads.html}}. The former contains about 17 thousand anime movies, and ratings from about 325 thousand users. The latter contains about 2 million books, and ratings from about 876 thousand users. Both datasets include item genres that we use as item features. 

We apply the following preprocessing procedure to prepare the datasets for the experiments. We remove items (and their ratings) whose genres are unknown. For the Anime dataset, we also remove movies whose titles are not written in the Roman alphabet and whose release date is from 2021 onwards to increase the chance that LLMs have been exposed to data about these movies. This is not necessary for the Goodreads dataset because the books all have titles in the Roman alphabet and were published before 2018. We then map ratings into a 1 to 5 scale and filter out users with fewer than 70 and more than 300 ratings. Table \ref{table:datasets_statistics} reports the statistics of the datasets after the preprocessing. 

\begin{table}[t]
\small
\caption{Statistics of the datasets}
\begin{tabular}{|c|c|c|c|c|c|c|c|c|}
\hline
          & \textbf{\# ratings} & \textbf{\# users} & \textbf{\# items} & \begin{tabular}[c]{@{}c@{}}\textbf{avg. \# ratings}\\ \textbf{per user}\end{tabular} & \begin{tabular}[c]{@{}c@{}}\textbf{avg. \# ratings}\\ \textbf{per item}\end{tabular} & \textbf{\# genres} & \begin{tabular}[c]{@{}c@{}}\textbf{avg. length of}\\ \textbf{item titles}\end{tabular} & \textbf{sparsity} \\ \hline
Anime     & 17M        & 118k     & 2.6k     & 143                                                                & 6.4k                                                               & 40        & 20.6                                                                & 94.50\%  \\ \hline
Goodreads & 8M         & 166k     & 8k       & 50                                                                 & 1k                                                                 & 16        & 37.1                                                                & 99.40\%  \\ \hline
\end{tabular}
\label{table:datasets_statistics}
\end{table}

\subsection{Methodology}\label{sec:methodology}

The following methodology applies to both datasets. We randomly split the dataset into training and test sets. We use a user-based split, with 80-20\% proportions. We train a Matrix Factorization baseline recommender (which we denote with MF) on the training set, using the RankSys implementation\footnote{\url{https://github.com/RankSys/RankSys}} of the factorization algorithm that is proposed in \cite{pilaszy2010fast}. We then randomly sample 500 test users and their data to run the experiments with the following procedure. 

For each test user, we use the MF baseline to compute a ranked candidate item list $CL$, such that none of the items in $CL$ is in the training dataset (i.e.\ MF recommends only unseen items). We employ InstructGPT (gpt-3.5-turbo-instruct version) and ChatGPT (gpt-3.5-turbo-0613 version) from the OpenAI GPT family\footnote{\url{https://platform.openai.com/docs/models/gpt-3-5}}, and Meta Llama2-Chat (7B\footnote{\url{https://huggingface.co/meta-llama/Llama-2-7b-chat-hf}} and 13B\footnote{\url{https://huggingface.co/meta-llama/Llama-2-13b-chat-hf}} versions) to re-rank $CL$ as outlined in Section \ref{sec:diversity_reranking}. We used the proprietary OpenAI API service when re-ranking with InstructGPT and ChatGPT.\footnote{\url{https://openai.com/index/openai-api/}. At the time we ran the experiments, the inference cost for InstructGPT was 1.5\$/1M for input tokens and 2\$/1M for output tokens; for ChatGPT, the inference cost was 0.5\$/1M for input tokens and 1.5\$/1M for output tokens.} In the case of the non-proprietary Llama2-7B-Chat and Llama2-13B-Chat, we used two NVIDIA A40 48GB GPUs to host the models and perform re-ranking.

As anticipated, LLMs can return partially invalid outputs. When the output is invalid in fairly simple ways, we try to remedy the situation intelligently. In cases when the LLM appends to the items' titles additional details (e.g.\ lists of genres, item descriptions, release dates), despite all prompts including clear instructions to outlaw this, we remove such extra content; we also remove any extra content that is prepended or appended to the final recommendation list (e.g.\ repetition of the re-ranking instructions from the prompt such as ``here is your re-ranked list of anime''); and we implement very simple inexact string matching methods (e.g.\ LLMs tend to add an extra space when encountering semicolons in item titles and our inexact matching takes care of this). But there are cases where invalid output is not so easily remedied, especially when the items themselves are invalid. Examples of invalid items include the following: the output list $RL$ may contain items whose titles did not appear in the candidate list $CL$; and there may be cases where there is a significant mismatch between the title of the item in $CL$ and its title in $RL$, e.g.\ ``\verb|Naruto|'' instead of ``\verb|Naruto:Shippuden|''. We discard these items from $RL$ because there is not a straightforward solution to address the invalid generation. In these cases, the $RL$ is incomplete, $|RL| < n$. We fill the list by inserting random items from $CL$ until $|RL| = n$. Instead, we could, but do not, repeat the generation step to try to get a valid list of items: at this stage, we prefer to quantify the problem and leave more advanced solutions for future work.  

Prompts T5-T8 presented in Section \ref{sec:diversity_reranking} need item features such as genres or item descriptions such as plot summaries. The former are already included in the datasets (see Table \ref{table:datasets_statistics}). The latter are extracted beforehand for each item by prompting ChatGPT with the following prompt: \verb|Please provide a one-sentence description of the following| \verb|item: {item name}|.

To compare our LLM-based re-rankers, we employ the MMR, xQuAD and RxQuAD greedy re-rankers presented in Section \ref{sec:increasing_div}. Their re-ranking strategy is explicit, they are fast and effective at diversifying recommendations, and they can use the item genres already available in the datasets as features to re-rank the $CL$ of each user. Note that, unlike the LLM-based re-rankers and the MRR re-ranker, xQuAD and RxQuAD additionally make use of the training data (to calculate aspect probabilities) to produce the re-ranking. In Section \ref{sec:results}, we show that this might explain why xQuAD and RxQuAD perform better than the other re-rankers on some metrics. We use the implementations of MRR, xQuAD and RxQuAD that are provided within the RankSys library, mentioned earlier. As a baseline, we also use a Random re-ranker from the same library that randomly samples items from $CL$ to generate $RL$. We point out that, although we could use other more complex (and expensive) re-rankers for comparison (e.g.\ \cite{wilhelm2018practical}), our focus is on exploring the applicability and the relative performance of LLM-based re-ranking, for which the MRR, xQuAD and RxQuAD re-rankers suffice.

\subsection{Hyperparameter tuning}\label{sec:hypers_tuning}
We use a validation set, a random sample of 20\% of the ratings from the training set, to tune the following hyperparameters for each dataset. We set the number $k$ of factors of the baseline recommender MF, choosing $k$ from $\{20, 50, 100, 150\}$ and optimizing for NDCG at cutoff 10. We found $k=20$ for the Anime dataset and $k=50$ for the Goodreads dataset. As is common, we set $n$, the size of the final set of recommendations to 10. Finally, to set $m$, first we find the mean $\mu$ and standard deviation $\sigma$ of the greatest rank in $CL$ of an item inserted in $RL$ for each re-ranker, i.e.\ $\max\limits_{\forall i \in RL}\{rank_{CL}(i)\}$; second, we take the largest value among these and set $m = \mu + \sigma$. We found $m=40$ for the Anime dataset and $m=50$ for the Goodreads dataset. This means that the values we are using for $m$ are bigger than those we find in the closest works to ours in the literature, e.g.\ in \cite{hou2024llm_rankers, sun2023chatgptgoodatsearch}, $m=20$; in \cite{dai2023uncovering}, $m=5$. This will make LLM-based re-ranking in our scenario more challenging because we will have greater prompt lengths and higher inference costs. But, at the same time, our scenario is more realistic because in real-world applications $m$ is unlikely to be as small as 5 or 20. 

\subsection{Evaluation metrics}\label{sec:eval_metrics}
To assess the relevance of recommendations, we measured Precision, Recall and NDCG at cutoff 10. However, in Section \ref{sec:results}, we report only NDCG because the three metrics reveal consistent findings. To measure recommendation diversity, we employ $\alpha$-NDCG, Intra-List Diversity (ILD), Expected Intra-List Diversity (EILD), Subtopic-Recall (SRecall) and relevancy-aware SRecall (rSRecall), extensively discussed in \cite{vargas2015thesis}. ILD calculates the average pairwise distance of the items in $RL$; to measure the distance, we use the complement of the Jaccard similarity between pairs of items  calculated on the item genres. SRecall calculates the proportion of genres covered by the items in $RL$. EILD and rSRecall calculate ILD and SRecall on the sublist of relevant items in $RL$, respectively. $\alpha$-NDCG is based on NDCG, but it is designed to account for relevance but also coverage of aspects (i.e.\ genres) and their redundancy; we set $\alpha=0.5$ following the argument in \cite{vargas2015thesis}.

\begin{table}[t]
\caption{Re-ranking performance for the Anime dataset. We report the scores achieved by the baseline MF on the six metrics. Then, for each re-ranker, we report the performances in terms of percentage difference with respect to the scores of the baseline MF. We report the average performance across the eight templates for each LLM-based re-ranker. All metrics are computed with a cutoff of size $n = |RL| =10 $. For all metrics, the higher the scores, the better the performance, and we highlight in bold the best performance for each metric.}
\begin{tabular}{c|c|c|c|c|c|c|c}
\cline{2-7}
                                         & \textbf{NDCG}  & \textbf{$\alpha$-NDCG} & \textbf{EILD} & \textbf{ILD}  & \textbf{rSRecall} & \textbf{SRecall} & \multicolumn{1}{l}{}                                                                            \\ \cline{1-7}
\multicolumn{1}{|c|}{MF}                 & \textbf{0.316} & 0.325            & 0.221         & 0.776         & 0.234             & 0.473            & \multicolumn{1}{l}{}                                                                            \\ \cline{1-7}
\multicolumn{1}{|c|}{\textbf{Re-ranker}} &                &                  &               &               &                   &                  & \multicolumn{1}{l}{}                                                                            \\ \hline
\multicolumn{1}{|c|}{Random}             & -58.8          & -52.6            & -57.3         & 4.5           & -38.0             & 5.6              & \multicolumn{1}{c|}{\multirow{8}{*}{\begin{tabular}[c]{@{}c@{}}\textbf{Difference}\\ \textbf{(\%)}\end{tabular}}} \\ \cline{1-7}
\multicolumn{1}{|c|}{MMR}                & -5.4           & -3.4             & \textbf{0.4}  & \textbf{12.6} & 1.5               & \textbf{15.0}    & \multicolumn{1}{c|}{}                                                                           \\ \cline{1-7}
\multicolumn{1}{|c|}{xQuad}              & -4.6           & \textbf{7.9}     & -4.2          & 0.5           & \textbf{7.5}      & 14.0             & \multicolumn{1}{c|}{}                                                                           \\ \cline{1-7}
\multicolumn{1}{|c|}{RxQuad}             & -6.5           & 4.1              & -6.4          & -0.8          & 4.3               & 7.0              & \multicolumn{1}{c|}{}                                                                           \\ \cline{1-7}
\multicolumn{1}{|c|}{ChatGPT}            & -10.1   & -6.0      & -9.4   & 4.7     & -2.5        & 7.8        & \multicolumn{1}{c|}{}                                                                           \\ \cline{1-7}
\multicolumn{1}{|c|}{InstructGPT}        & -14.9  & -10.2     & -14.7 & 4.7     & -6.1       & 7.2        & \multicolumn{1}{c|}{}                                                                           \\ \cline{1-7}
\multicolumn{1}{|c|}{Llama2-7B-Chat}     & -21.5  & -15.5    & -23.3 & 4.6     & -10.8      & 8.0        & \multicolumn{1}{c|}{}                                                                           \\ \cline{1-7}
\multicolumn{1}{|c|}{Llama2-13B-Chat}    & -26.0  & -21.6    & -25.1 & 4.4     & -14.7       & 4.7        & \multicolumn{1}{c|}{}                                                                           \\ \hline
\end{tabular}
\label{table:anime_results_traditional}
\end{table}

\section{Results}\label{sec:results}

For ease of exposition, we divide this section into three subsections. In the first, we focus our analysis on comparing LLM-based re-rankers with a random re-ranker (we want to know whether LLM re-ranking is more than arbitrary) and traditional re-rankers (we want to know how LLM re-ranking is positioned with respect to simple state-of-the-art approaches). In the second subsection, we focus our analysis on comparing different prompt templates for LLM-based re-rankers: we want to know how different prompts affect performance. Lastly, in the third subsection we analyse the costs incurred by the different re-rankers. 

Tables \ref{table:anime_full_results} and \ref{table:goodreads_full_results} in Appendix B report the full set of results presented in this section. In these tables we also report confidence intervals calculated with a 95\% confidence level and we use them through our analysis to help outline the uncertainty of our results. Non-overlapping intervals imply statistical significance between performances while, despite common belief, overlapping intervals do not necessarily imply lack of statistical significance \cite{greenland2016statistical}. In fact, the confidence intervals in our results are quite large for all metrics except for ILD and SRecall --- due to high variability in the values of the metrics but also the small number of users in our experiments, i.e.\ 500. 
This limits the number of cases where we can be sure of statistical significance (non-overlapping intervals). In future work,  we hope to extend our experiments with a larger group of users and that will allow us to perform a deeper statistical analysis 
and strengthen our findings. This should be possible assuming that the costs (both monetary and computational) of using LLMs continues to fall.

\subsection{Comparison of LLM-based re-rankers with random and traditional re-rankers}

\begin{table}[t]
\caption{Re-ranking performance for the Goodreads dataset. We report the scores achieved by the baseline MF on the six metrics. Then, for each re-ranker, we report the performances in terms of percentage difference with respect to the scores of the baseline MF. We report the average performance across the eight templates for each LLM-based re-ranker. All metrics are computed with a cutoff of size $n = |RL| =10 $. For all metrics, the higher the scores, the better the performance, and we highlight in bold the best performance for each metric.}
\begin{tabular}{c|c|c|c|c|c|c|c}
\cline{2-7}
                                         & \textbf{NDCG}  & \textbf{$\alpha$-NDCG} & \textbf{EILD} & \textbf{ILD}  & \textbf{rSRecall} & \textbf{SRecall} & \multicolumn{1}{l}{}                                                                            \\ \cline{1-7}
\multicolumn{1}{|c|}{MF}                 & \textbf{0.249} & 0.271            & 0.092         & 0.689         & 0.355             & 0.871            & \multicolumn{1}{l}{}                                                                            \\ \cline{1-7}
\multicolumn{1}{|c|}{\textbf{Re-ranker}} &                &                  &               &               &                   &                  & \multicolumn{1}{l}{}                                                                            \\ \hline
\multicolumn{1}{|c|}{Random}             & -66.3          & -59.6            & -71.3         & 2.0           & -43.8             & 2.1              & \multicolumn{1}{c|}{\multirow{8}{*}{\begin{tabular}[c]{@{}c@{}}\textbf{Difference}\\ \textbf{(\%)}\end{tabular}}} \\ \cline{1-7}
\multicolumn{1}{|c|}{MMR}                & -4.7           & -6.6             & \textbf{-2.5} & \textbf{17.7} & -5.3              & 7.2              & \multicolumn{1}{c|}{}                                                                           \\ \cline{1-7}
\multicolumn{1}{|c|}{xQuad}              & -7.8           & \textbf{13.7}    & -9.3          & -8.2          & \textbf{8.8}      & \textbf{10.0}    & \multicolumn{1}{c|}{}                                                                           \\ \cline{1-7}
\multicolumn{1}{|c|}{RxQuad}             & -8.1           & 11.9             & -8.6          & -7.7          & 7.4               & 8.6              & \multicolumn{1}{c|}{}                                                                           \\ \cline{1-7}
\multicolumn{1}{|c|}{ChatGPT}            & -33.4  & -19.3    & -46.0 & -4.8   & -20.0     & 1.8        & \multicolumn{1}{c|}{}                                                                           \\ \cline{1-7}
\multicolumn{1}{|c|}{InstructGPT}        & -53.5  & -36.6    & -66.5 & -4.7   & -35.0     & 1.4        & \multicolumn{1}{c|}{}                                                                           \\ \cline{1-7}
\multicolumn{1}{|c|}{Llama2-7B}     & -59.5  & -50.4    & -70.0 & -1.1   & -43.7     & 1.1        & \multicolumn{1}{c|}{}                                                                           \\ \cline{1-7}
\multicolumn{1}{|c|}{Llama2-13B}    & -61.3  & -48.4    & -71.9 & -2.2   & -41.8     & 1.5        & \multicolumn{1}{c|}{}                                                                           \\ \hline
\end{tabular}
\label{table:goodreads_results_traditional}
\end{table}

Tables \ref{table:anime_results_traditional} and \ref{table:goodreads_results_traditional} report the results of the re-rankers on the Anime and Goodreads datasets, respectively. For each re-ranker, we report the percentage difference with respect to the baseline recommendations from MF. For the LLM-based re-rankers, we report the average performance across the eight templates. 
We postpone the fine-grained analysis across different prompt templates to the next subsection. 

As expected, re-ranking decreases relevance and increases the diversity of recommendations for both datasets. In particular, for the Random re-ranker, relevance, measured by NDCG, drops significantly, between 58.8\% and 66.3\%. Relevance also plays a major role in the $\alpha$-NDCG, EILD and rSRecall metrics, so we see large decreases in their values for the Random re-ranker. The diversity of its outputs (ILD and SRecall) increases, as typically happens when introducing randomness into recommendations, but only slightly for Goodreads. 

In both datasets, all the intelligent re-rankers (both the traditional ones and the LLM-based ones) outperform the Random re-ranker for all the metrics to a great extent (and their performances are statistically significantly different), except for ILD and SRecall (where the superiority is not consistent across all intelligent re-rankers). This confirms what we were looking to prove with RQ2: that LLMs can interpret diversification re-ranking and produce rankings that are more than arbitrary.

Regarding traditional re-ranking, MMR and xQuAD seem to have a better trade-off than RxQuAD, but this strongly depends on which metrics one considers to be more important. RxQuAD has the highest drop in relevance as measured by NDCG in both datasets and, while xQuAD is superior in this metric to MMR for the Anime dataset, the opposite stands for the Goodreads dataset. As expected, MMR is the best for ILD on both datasets since it optimizes this metric. While MMR also improves SRecall on both datasets (even having the best score on the Anime dataset), it fails to improve $\alpha$-NDCG, and it slightly improves the other relevance-aware diversity metrics (i.e.\ EILD and rSRecall) on the Anime dataset but not in the Goodreads dataset.

On both datasets, xQuAD is the best re-ranker at improving $\alpha$-NDCG as this metric is very similar to what it optimises. Similarly, also RxQuAD improves on this metric as expected, but to a lesser extent than xQuAD. xQuAD improves the other diversity metrics too, except for EILD and ILD on both datasets (and it is the best for rSRecall and SRecall on both datasets and for SRecall on Goodreads). Finally, RxQuAD improves $\alpha$-NDCG, rSRecall and SRecall (as expected, being similar to xQuAD), but in none of these metrics does its scores outperform all the other traditional re-rankers. The lower performance of RxQuAD in terms of relevance-aware metrics is quite surprising since this approach is designed to focus on relevance more than the other two re-rankers. Other works have found similar outcomes \cite{kaya2019subprofile}: we posit that this is because the datasets exhibit a weak correlation between relevance and item features, in contrast with the assumption made by the re-ranking objective function RxQuAD is built upon.

\begin{table}[t]
\caption{Average lowest rank of items in the candidate list $CL$ recommended in the final list $RL$. High values correspond to low ranks. To calculate the lowest rank, we do not consider the random recommendations in $RL$ provided to the user when invalid generations occur.}
\centering
\begin{tabular}{|c|c|c|}
\hline
\textbf{Re-ranker}               & \textbf{Anime} & \textbf{Goodreads} \\ \hline
MMR            & 19.5  & 22.58     \\ \hline
xQuAD          & 21    & 17.9      \\ \hline
RxQuAD         & 16    & 15.9      \\ \hline
ChatGPT        & 22.6  & 31.9      \\ \hline
InstructGPT    & 22.3  & 37.7      \\ \hline
Llama2-7B-Chat & 25.5  & 39.5      \\ \hline
Llama2-13B-Chat & 28.9  & 43.1      \\ \hline
\end{tabular}
\label{table:avg_max_rank}
\end{table}

\begin{table}[t]
\small
\caption{Average percentage of random recommendations (due to invalid generation) provided to the 500 users, distributed across the datasets, the LLM-based re-rankers and the prompt templates. In bold we report the lowest value in each group.}
\begin{tabular}{|cc|ccc|cccccccc|}
\hline
\multicolumn{2}{|c|}{\textbf{Dataset}} & \multicolumn{3}{c|}{\textbf{Re-ranker}}                & \multicolumn{8}{c|}{\textbf{Prompt template}}                                              \\ \hline
Anime            & Goodreads  & ChatGPT & InstructGPT    & Llama2-7B-Chat & T1    & T2             & T3    & T4    & T5    & T6    & T7     & T8     \\ \hline
\textbf{2.3\%}   & 13.1\%     & 3.0\%   & \textbf{1.2\%} & 18.9\%         & 4.0\% & \textbf{3.1\%} & 8.8\% & 4.8\% & 9.3\% & 8.9\% & 18.0\% & 22.4\% \\ \hline
\end{tabular}
\label{table:null_recs_prop}
\end{table}

On the Anime dataset, the decrease in NDCG for the LLM-based re-rankers (roughly 10-26\% loss) is greater than that for the traditional re-rankers (4.6-6.5\%), although  at its best (i.e.\ for the template on which it performs best) ChatGPT outperforms all the other re-rankers (losing just 2.8\%); see the full results in the Appendix. There are similar results for the other metrics, particularly those where recommendation relevance plays a major role (i.e.\ $\alpha$-NDCG, EILD and rSRecall). In particular, traditional re-rankers are statistically significantly better than LLM-based re-rankers (except ChatGPT) for NDCG, $\alpha$-NDCG and EILD. The metric where LLM-based re-ranking does best in comparison with traditional re-ranking is ILD (and, where better, statistical significance is achieved). For this metric, the OpenAI re-rankers perform better than xQuAD and RxQuAD. Finally, there is a clear difference between the LLM-based re-rankers themselves: ChatGPT is superior to all others according to all metrics (except for SRecall where Llama2-7B-Chat is the best). Also, OpenAI models are better than Meta ones, with Llama2-13B-Chat being the worst of all the re-rankers.

On the Goodreads dataset, there is a similar pattern of results for the LLM-based re-ranking, but this time the performance gap with traditional re-rankers is higher across all metrics (in particular for NDCG and $\alpha$-NDCG). ChatGPT is not as close to some traditional re-rankers as it was on the Anime dataset (and differences are statistically significant). It is still the case that the LLM-based re-rankers perform better than xQuAD and RxQuAD for ILD (with a statistically significant difference), and, among themselves, their performance from best to worst remains ChatGPT, then InstructGPT, then Llama2-7B and lastly Llama2-13B.

Three factors might explain why LLM-based re-rankers are inferior to traditional re-rankers in these experiments, particularly for relevance-aware metrics. First, we argue that relevance in $RL$ should fall when drawing low-rank items from $CL$, assuming that items in $CL$ are accurately ranked by their relevance. Indeed, Table \ref{table:avg_max_rank} shows that LLM-based re-rankers draw items ranked lower in $CL$ than the ones drawn by traditional re-rankers. Second, the proportion of random recommendations in the final $RL$ lists of LLM-based re-rankers (Table \ref{table:null_recs_prop}) might harm relevance, in particular for the Goodreads dataset. (As we explained in Section \ref{sec:methodology}, when the output of the LLM contains invalid items, we replace the invalid items with randomly chosen items from $CL$, and this is likely to harm the relevance of the final $RL$.) Third, traditional re-rankers such as xQuAD and RxQuAD might have an advantage over LLM-based re-ranking as the former explicitly use user profiles to estimate relevance and genre distributions within profiles, while the latter do not have access to such information. Indeed, the LLMs might struggle to capture relevance (in terms of rank) and diversity (in terms of genres) and the interplay between the two from the prompt templates we have designed, even though the prompts explicitly highlight item relevance (by including each item's rank) and some of them (\verb|T5|--\verb|T8|) highlight diversity (by providing an item's genres or a description of the item) --- see the detailed analysis of the prompt template results in the next subsection.

\begin{figure}[t]
  \centering
  \includegraphics[width=0.98\textwidth]{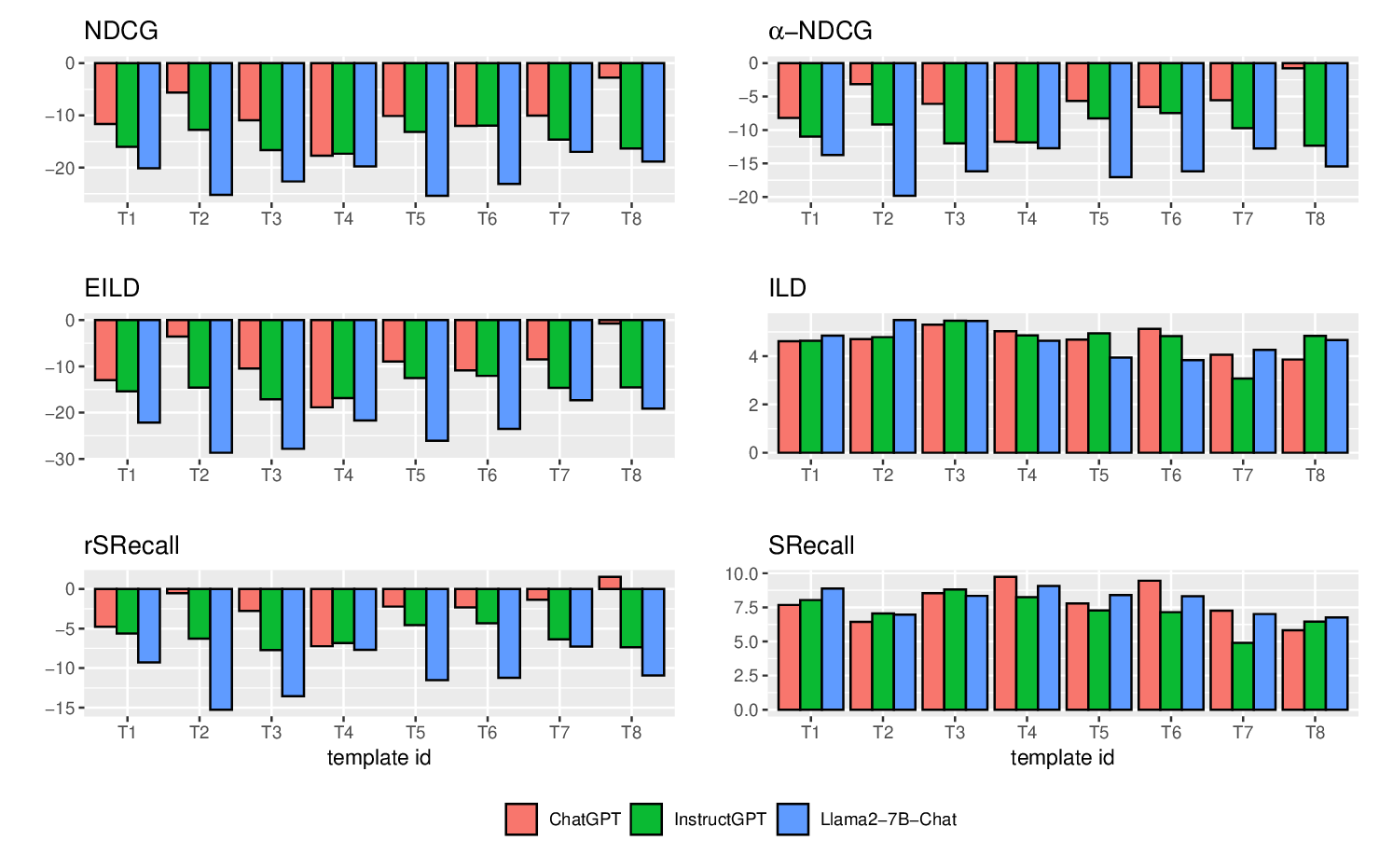}
  \caption{Comparison of the LLM-based re-rankers across different prompt templates for the Anime dataset. For each metric, we plot the percentage difference with respect to the scores of the baseline MF.}
\label{fig:anime_results_plot}
\end{figure}

\subsection{Comparison LLM-based re-rankers}

\begin{figure}
  \centering
  \includegraphics[width=0.98\textwidth]{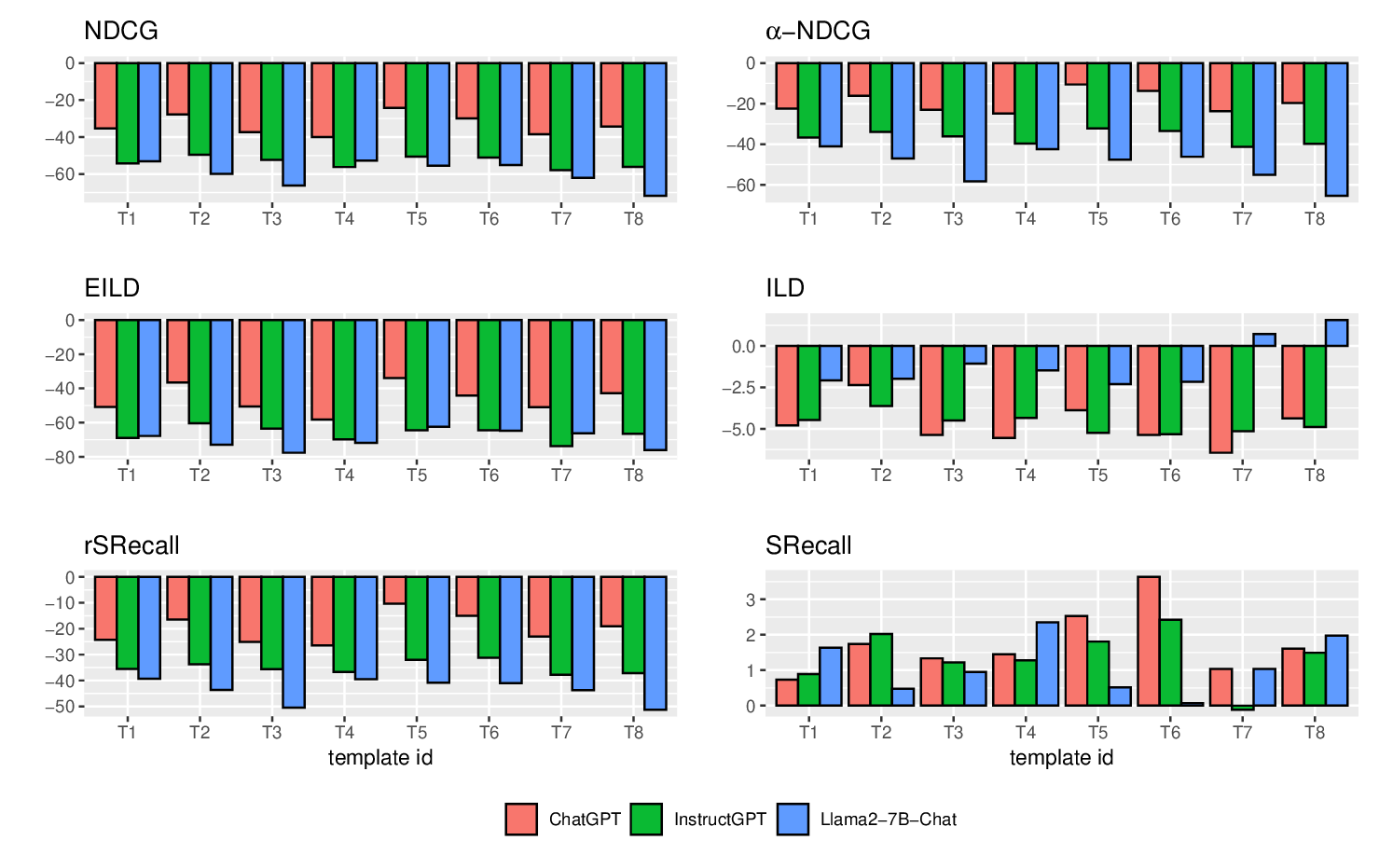}
  \caption{Comparison of the LLM-based re-rankers across different prompt templates for the Goodreads dataset. For each metric, we plot the percentage difference with respect to the scores of the baseline MF.}
  \label{fig:goodreads_results_plot}
\end{figure}

We use Figures \ref{fig:anime_results_plot} \& \ref{fig:goodreads_results_plot} and Table 
\ref{table:templates_performance} to draw a comparison across different prompt templates and LLM-based re-rankers for both the Anime and Goodreads datasets. In Figures \ref{fig:anime_results_plot} and \ref{fig:goodreads_results_plot} for Anime and Goodreads dataset respectively, each plot corresponds to a different metric. In each plot, on the horizontal axis, there are groups of results for each of the eight templates (T1--T8) and, within each group, there are three re-rankers, i.e.\ ChatGPT, InstructGPT, Llama2-7B-Chat. On the vertical axis, values represent the percentage difference between the re-ranker and the MF baseline. 
Table \ref{table:templates_performance} reports the performances of the LLM-based re-rankers grouped by prompt template.

For ease of exposition, we do not include results for Llama2-13B-Chat (although they, along with all results are in the Appendix). Its performances is generally worse than all the other re-rankers, hence its results do not add valuable insights to the discussion.  It is surprising that Llama2-13B-Chat is worse than Llama2-7B-Chat, since LLM capabilities usually correlate with LLM size when considering models of the same family. However, differences in human preference alignment may play a major confounding role \cite{ouyang2022instructgptalignment}, and there are indications in the literature that Llama2-13B-Chat is worse than Llama2-13B-Chat for some tasks, e.g.\ see human evaluation in \cite{touvron2023llama}. But, beyond these general points, we have no clear explanation for why Llama2-13B-Chat performs so poorly in our experiments and we leave its investigation for future work.

Overall, we can draw three main findings. First, LLM-based re-ranking improves the baseline ranking of MF for ILD in the Anime dataset and SRecall in both datasets (and all the LLM-based re-ranking performances except for the ones of ChatGPT are statistically significantly different from the MF ones). For the other metrics, performance degradation is more severe on the Goodreads than on the Anime dataset. The fact that the Anime dataset is less prone to invalid generation than the Goodreads dataset (2.3\% vs 13.1\%, Table \ref{table:null_recs_prop}) might partially explain the difference in degradation. Also, an LLM's internal knowledge about the specific recommendation domain is another key factor affecting re-ranking. We speculate that generating content related to books is more challenging than anime, probably due to the nature of the domain and its content (e.g.\ the presence of similar titles in a prompt or the length of the prompts, which are longer for books than for anime on average, see Table \ref{table:datasets_statistics}). However, our experiments cannot confirm our speculation or that this even applies to other domains. We plan to address this in future work.

Second, overall, ChatGPT exhibits the best performance among LLMs (within each specific template and averaged across all templates) for all metrics, with the only exception being SRecall and ILD in both datasets, where the model is not the best in all the different templates. We posit that the extent to which an LLM re-ranker draws low-rank items and includes random recommendations from $CL$ in the final recommendation list $RL$ correlates with performance differences across datasets and templates, which partially explain these results. Indeed, Tables \ref{table:avg_max_rank} and \ref{table:null_recs_prop} show that OpenAI's models (on average) draw higher ranked items from $CL$ and need fewer random recommendations than Meta's models. This is consistent across the Anime and Goodreads datasets, but to a greater extent for the former. The only exception is between ChatGPT and InstructGPT, where the former has more random recommendations than the latter. Thus, we argue that the proportion of random recommendations only partly explains the loss of relevance in LLM-based re-rankers. We believe that a model's reasoning capability also plays a major role in performance differences. According to several benchmarks\footnote{https://artificialanalysis.ai/models} and the literature \cite{chiang2024chatbotarena, kaur2023evaluating},  OpenAI models are superior to Meta's models, and ChatGPT shows superior capabilities over InstructGPT  (although these benchmarks and the literature are not specific to recommender systems). Our results on diversification re-ranking seem to confirm what the benchmarks and literature have shown.

Third, different prompt templates show different performance across re-rankers, datasets and metrics. This suggests that there is no best template. Prompt template design needs to be tailored to specific LLMs, recommendation domains and metrics. 
In the next two paragraphs, we will highlight the most significant of our findings about the different templates. 

Templates \verb|T1| and \verb|T4|--\verb|T7| are designed to strike a balance between relevance and diversity (similar to setting $\lambda=0.5$) and should exhibit the best performance for relevancy-aware metrics, while \verb|T2|, \verb|T3| and \verb|T8| should increase diversity the most ($\lambda=0$). This is reflected in their average re-ranking performances, but not consistently (see Table \ref{table:templates_performance}). On the Anime dataset, for example, \verb|T8| and \verb|T7| are the templates where relevance-aware metrics degrade the least, even though \verb|T8| is not explicitly designed to account for relevance. Also, \verb|T2|, \verb|T3| and \verb|T4| improve ILD and rSRecall the most, even though \verb|T4| should focus more on balancing relevance (but instead degrades relevance-aware metrics more than any other templates). The results of the Goodreads dataset are more consistent with what we expect from the template design. \verb|T5| and \verb|T6| show the least amount of degradation on the relevance-aware metrics as expected;\verb|T2|, \verb|T8| benefit ILD the most; and \verb|T4|, \verb|T6| \& \verb|T8| benefit SRecall the most.

On average, feature-aware templates (i.e.\ \verb|T5|--\verb|T8|) benefit the re-ranking performances more than templates that do not include item features (i.e.\ \verb|T5|--\verb|T8|) on both datasets, even though the feature-aware templates result in more invalid and hence random recommendations (which may affect performance negatively). Notably, \verb|T8| has the best performance on relevance-aware metrics for the Anime dataset and on diversity metrics for the Goodreads dataset despite generating the largest number of invalid recommendations (22.8\%). Overall, these findings are a positive indication that LLMs can effectively use additional information to improve re-ranking results.

\begin{table}[t]
\small
\caption{Average performances of ChatGPT, InstructGPT and Llama2-7B-Chat with different prompt templates. For each metric, we highlight the two best performances in bold.}
\begin{tabular}{|c|cccccc|cccccc|}
\hline
\multicolumn{1}{|l|}{} & \multicolumn{6}{c|}{\textbf{Anime}}                                                                                                                                                                               & \multicolumn{6}{c|}{\textbf{Goodreads}}                                                                                                                                                                            \\ \hline
\multicolumn{1}{|l|}{} & \multicolumn{1}{c|}{\textbf{NDCG}}  & \multicolumn{1}{c|}{\textbf{$\alpha$-NDCG}} & \multicolumn{1}{c|}{\textbf{EILD}}  & \multicolumn{1}{c|}{\textbf{ILD}} & \multicolumn{1}{c|}{\textbf{rSRecall}} & \textbf{SRecall} & \multicolumn{1}{c|}{\textbf{NDCG}}  & \multicolumn{1}{c|}{\textbf{$\alpha$-NDCG}} & \multicolumn{1}{c|}{\textbf{EILD}}  & \multicolumn{1}{c|}{\textbf{ILD}}  & \multicolumn{1}{c|}{\textbf{rSRecall}} & \textbf{SRecall} \\ \hline
\textbf{T1}            & \multicolumn{1}{c|}{-15.9}          & \multicolumn{1}{c|}{-11.0}            & \multicolumn{1}{c|}{-16.9}          & \multicolumn{1}{c|}{4.7}          & \multicolumn{1}{c|}{-6.6}              & 8.2              & \multicolumn{1}{c|}{-47.6}          & \multicolumn{1}{c|}{-33.4}            & \multicolumn{1}{c|}{-62.6}          & \multicolumn{1}{c|}{-3.8}          & \multicolumn{1}{c|}{-33.1}             & 1.1              \\ \hline
\textbf{T2}            & \multicolumn{1}{c|}{-14.5}          & \multicolumn{1}{c|}{-10.7}            & \multicolumn{1}{c|}{-15.6}          & \multicolumn{1}{c|}{\textbf{5.0}} & \multicolumn{1}{c|}{-7.3}              & 6.8              & \multicolumn{1}{c|}{-45.7}          & \multicolumn{1}{c|}{-32.4}            & \multicolumn{1}{c|}{\textbf{-56.6}} & \multicolumn{1}{c|}{\textbf{-2.7}} & \multicolumn{1}{c|}{-31.3}             & 1.4              \\ \hline
\textbf{T3}            & \multicolumn{1}{c|}{-16.7}          & \multicolumn{1}{c|}{-11.4}            & \multicolumn{1}{c|}{-18.5}          & \multicolumn{1}{c|}{\textbf{5.4}} & \multicolumn{1}{c|}{-8.0}              & \textbf{8.6}     & \multicolumn{1}{c|}{-52.0}          & \multicolumn{1}{c|}{-39.1}            & \multicolumn{1}{c|}{-64.0}          & \multicolumn{1}{c|}{-3.6}          & \multicolumn{1}{c|}{-37.1}             & 1.2              \\ \hline
\textbf{T4}            & \multicolumn{1}{c|}{-18.3}          & \multicolumn{1}{c|}{-12.1}            & \multicolumn{1}{c|}{-19.1}          & \multicolumn{1}{c|}{4.8}          & \multicolumn{1}{c|}{-7.2}              & \textbf{9.0}     & \multicolumn{1}{c|}{-49.6}          & \multicolumn{1}{c|}{-35.6}            & \multicolumn{1}{c|}{-66.6}          & \multicolumn{1}{c|}{-3.8}          & \multicolumn{1}{c|}{-34.2}             & \textbf{1.7}     \\ \hline
\textbf{T5}            & \multicolumn{1}{c|}{-16.2}          & \multicolumn{1}{c|}{-10.3}            & \multicolumn{1}{c|}{-15.9}          & \multicolumn{1}{c|}{4.5}          & \multicolumn{1}{c|}{-6.1}              & 7.8              & \multicolumn{1}{c|}{\textbf{-43.5}} & \multicolumn{1}{c|}{\textbf{-30.1}}   & \multicolumn{1}{c|}{\textbf{-53.6}} & \multicolumn{1}{c|}{-3.8}          & \multicolumn{1}{c|}{\textbf{-27.7}}    & 1.6              \\ \hline
\textbf{T6}            & \multicolumn{1}{c|}{-15.7}          & \multicolumn{1}{c|}{-10.1}            & \multicolumn{1}{c|}{-15.5}          & \multicolumn{1}{c|}{4.6}          & \multicolumn{1}{c|}{-6.0}              & 8.3              & \multicolumn{1}{c|}{\textbf{-45.3}} & \multicolumn{1}{c|}{\textbf{-31.1}}   & \multicolumn{1}{c|}{-57.8}          & \multicolumn{1}{c|}{-4.3}          & \multicolumn{1}{c|}{\textbf{-29.1}}    & \textbf{2.0}     \\ \hline
\textbf{T7}            & \multicolumn{1}{c|}{\textbf{-13.9}} & \multicolumn{1}{c|}{\textbf{-9.3}}    & \multicolumn{1}{c|}{\textbf{-13.5}} & \multicolumn{1}{c|}{3.8}          & \multicolumn{1}{c|}{\textbf{-5.0}}     & 6.4              & \multicolumn{1}{c|}{-52.8}          & \multicolumn{1}{c|}{-40.0}            & \multicolumn{1}{c|}{-63.7}          & \multicolumn{1}{c|}{-3.6}          & \multicolumn{1}{c|}{-34.9}             & 0.7              \\ \hline
\textbf{T8}            & \multicolumn{1}{c|}{\textbf{-12.7}} & \multicolumn{1}{c|}{\textbf{-9.5}}    & \multicolumn{1}{c|}{\textbf{-11.5}} & \multicolumn{1}{c|}{4.5}          & \multicolumn{1}{c|}{\textbf{-5.6}}     & 6.3              & \multicolumn{1}{c|}{-54.1}          & \multicolumn{1}{c|}{-41.6}            & \multicolumn{1}{c|}{-61.8}          & \multicolumn{1}{c|}{\textbf{-2.6}} & \multicolumn{1}{c|}{-35.8}             & \textbf{1.7}     \\ \hline
\end{tabular}
\label{table:templates_performance}
\end{table}

\subsection{Re-ranking cost analysis}\label{sec:trade_off_cost}

Finally, we analyse the cost of performing re-ranking. Table \ref{table:breakdown_inference_time} reports the average inference time to re-rank the candidate recommendations of a user. For re-ranking using OpenAI's LLMs, the table shows the same value (i.e.\ 25 seconds). This value is not the actual cost. It is an upper bound which we have to use in this table due to problems we encountered using OpenAI APIs. In a preliminary experiment to test the APIs, we noticed that the model's response time had a very large variability and sometimes the model did not return any outputs at all (i.e.\ the service seemed not to be available), probably due to unknown usage policies and high workloads. We found that introducing a sleep time of 25 seconds between API calls was a way of ensuring we obtaining outputs. Unfortunately, this prevents us from providing an accurate estimation of response time. 
Nevertheless, the table shows that traditional re-rankers have a clear advantage over LLM-based re-rankers here: they are three orders of magnitude faster than the LLM-based re-rankers. In the case of LLM-based re-rankers, it also shows, as expected, that larger models lead to longer re-ranking time: Llama2-7B-Chat is faster than Llama2-13B-Chat.

\begin{table}[t]
\caption{Average computational time (in seconds) to re-rank the candidate recommendation of a single user in our experiments.}
\begin{tabular}{c|cccccl|}
\cline{2-7}
                                & \multicolumn{6}{c|}{\textbf{Re-rankers}}                                                                                                                                                                                                       \\ \cline{2-7} 
                                & \multicolumn{1}{c|}{\textbf{MMR}}   & \multicolumn{1}{c|}{\textbf{xQuAD}} & \multicolumn{1}{c|}{\textbf{RxQuAD}} & \multicolumn{1}{c|}{\textbf{ChatGPT/InstructGPT}} & \multicolumn{1}{c|}{\textbf{Llama2-7B-Chat}} & \textbf{Llama2-13B-Chat} \\ \hline
\multicolumn{1}{|c|}{Anime}     & \multicolumn{1}{c|}{0.05}           & \multicolumn{1}{c|}{0.1}            & \multicolumn{1}{c|}{0.122}           & \multicolumn{1}{c|}{\textless{}25}                & \multicolumn{1}{c|}{12.6}                    & \multicolumn{1}{c|}{23.7}                        \\ \hline
\multicolumn{1}{|c|}{Goodreads} & \multicolumn{1}{c|}{0.08}           & \multicolumn{1}{c|}{0.13}           & \multicolumn{1}{c|}{0.18}            & \multicolumn{1}{c|}{\textless{}25}                & \multicolumn{1}{c|}{16.1}                    & \multicolumn{1}{c|}{31.6}                        \\ \hline
                                & \multicolumn{1}{c|}{\textbf{0.065}} & \multicolumn{1}{c|}{0.115}          & \multicolumn{1}{c|}{0.151}           & \multicolumn{1}{c|}{\textless{}25}                & \multicolumn{1}{c|}{14.35}                   & \multicolumn{1}{c|}{27.6}                        \\ \cline{2-7} 
\end{tabular}
\label{table:breakdown_inference_time}
\end{table}

Table \ref{table:breakdown_costs_openai} breaks down the financial costs in terms of API calls for the OpenAI models\footnote{The prices relate to the timeframe of our experiments, i.e.\ from October 2023 to January 2024.Prcies and pricing policies may have changed.}. There are no API costs for re-ranking using Meta's LLMs since we hosted these models on our own GPUs. In principle, we could provide a cost comparison between the OpenAI and Meta model re-rankers if we were to calculate their energy costs. In practice, estimates of energy costs are hard to give due, e.g., to the volatility of energy market prices over the period during which we ran the experiments. The total cost of re-ranking for 500 users in our experiment is roughly 161\$, and around 75\% of this expense is incurred by InstructGPT, whose API calls are more expensive. Prices are correlated with the number of tokens processed: the Anime dataset experiment is cheaper than the one with Goodreads as the latter's item titles and features are more wordy; feature-aware prompt templates (i.e.\ T5-T8) are the most expensive ones (in particular T7 and T8, because item descriptions in the form of  plot summaries generally have more tokens than genre lists). Therefore, recalling the results of the earlier sections, we find that ChatGPT is superior to InstructGPT in terms of accuracy/diversity and cost.

In summary, traditional re-ranking is much faster and less resource-demanding than LLM-based re-ranking. Among LLM-based re-ranking, ChatGPT is cheaper than InstructGPT, and use of these models comes with a lack of control and transparency due to the API services they rely on. Finally, Meta's models ensure more control over data and inference but need private computational resources, with difficult-to-estimate costs that are subject to market volatility. However, some recent reports\footnote{https://www.ark-invest.com/big-ideas-2024} report that training and inference are rapidly becoming cheaper (decreasing annually by 75\% and ~86\% respectively), which will probably break down the barrier for LLM adoption and reduce the cost advantage of traditional re-ranking.

\begin{table}[t]
\caption{Total re-ranking costs using the OpenAI models for the 500 users in our experiments.}
\begin{tabular}{c|ccccc|}
\hline
\multicolumn{1}{|c|}{\multirow{2}{*}{\textbf{Dataset}}}   & \multicolumn{1}{c|}{\multirow{2}{*}{\textbf{Template id}}} & \multicolumn{2}{c|}{\textbf{\# Input tokens}}                             & \multicolumn{2}{c|}{\textbf{\# Output tokens}}       \\ \cline{3-6} 
\multicolumn{1}{|c|}{}                           & \multicolumn{1}{c|}{}                             & \multicolumn{1}{c|}{\textbf{ChatGPT}}  & \multicolumn{1}{c|}{\textbf{InstructGPT}} & \multicolumn{1}{c|}{\textbf{ChatGPT}}  & \textbf{InstructGPT} \\ \hline
\multicolumn{1}{|c|}{\multirow{3}{*}{Anime}}     & \multicolumn{1}{c|}{T1-T4}                        & \multicolumn{2}{c|}{$\sim$1.8M}                                   & \multicolumn{1}{c|}{$\sim$118k}     & $\sim$121k        \\
\multicolumn{1}{|c|}{}                           & \multicolumn{1}{c|}{T5, T6}                       & \multicolumn{2}{c|}{$\sim$4.2M}                                   & \multicolumn{1}{c|}{$\sim$159k}     & $\sim$210k        \\
\multicolumn{1}{|c|}{}                           & \multicolumn{1}{c|}{T7, T8}                       & \multicolumn{2}{c|}{$\sim$8.5M}                                   & \multicolumn{1}{c|}{$\sim$176k}     & $\sim$245k        \\ \hline
\multicolumn{1}{|c|}{\multirow{3}{*}{Goodreads}} & \multicolumn{1}{c|}{T1-T4}                        & \multicolumn{2}{c|}{$\sim$3.3M}                                   & \multicolumn{1}{c|}{$\sim$212k}     & $\sim$220k        \\
\multicolumn{1}{|c|}{}                           & \multicolumn{1}{c|}{T5, T6}                       & \multicolumn{2}{c|}{$\sim$6.9M}                                   & \multicolumn{1}{c|}{$\sim$221k}     & $\sim$285k        \\
\multicolumn{1}{|c|}{}                           & \multicolumn{1}{c|}{T7, T8}                       & \multicolumn{2}{c|}{$\sim$9.4M}                                   & \multicolumn{1}{c|}{$\sim$261k}     & $\sim$269k        \\ \hline
                                                 & \multicolumn{1}{c|}{Total \# of tokens}           & \multicolumn{2}{c|}{$\sim$78.4M}                                       & \multicolumn{1}{c|}{$\sim$1.1M}     & $\sim$1.3M        \\ \cline{2-6} 
                                                 & \multicolumn{1}{c|}{Unitary price}                & \multicolumn{1}{c|}{0.5\$/1M} & \multicolumn{1}{c|}{1.5\$/1M}    & \multicolumn{1}{c|}{1.5\$/1M} & 2\$/1M      \\ \cline{2-6}
                                                 & Total cost ChatGPT                                       & \multicolumn{4}{c|}{78.4M $\cdot$ 0.5 + 1.1M $\cdot$ 1.5\$ = \textbf{40.85\$}}                                     \\ \cline{2-6}
                                                 & Total cost InstructGPT                                       & \multicolumn{4}{c|}{78.4M $\cdot$ 1.5\$ + 1.3M $\cdot$ 2\$ = \textbf{120.2\$}}                                     \\ \cline{2-6}
                                                 & Total cost                                       & \multicolumn{4}{c|}{40.85\$ + 120.2\$ = \textbf{161.05\$}}                                     \\ \cline{2-6} 
\end{tabular}
\label{table:breakdown_costs_openai}
\end{table}

\section{Beyond performance: LLM-based and greedy re-ranking trade-offs}\label{sec:beyond_performance}
In the previous section we have analysed and discussed accuracy/diversity performance and costs of LLM-based re-rankers and compared them with the ones of traditional greedy re-ranking. In this section we focus on RQ4: we discuss the trade-offs RS providers should take into account when selecting one of these two re-ranking approaches. Table \ref{table:comparison_tradeoff} summarizes our analysis.


Many of the traditional approaches need explicit item features to use in the re-ranking. Acquiring these features might require human labelling or other expensive information retrieval procedures. The LLM-based approaches do not need to be given explicit item features to perform the re-ranking. They can instead take advantage of implicit knowledge about items that is embedded in their parameters during pre-training or human alignment. This implicit knowledge might be more complex than the simple feature sets (e.g.\ genres) typically used in traditional re-ranking, and this might enhance the diversity of the final recommendations. Moreover, LLMs might even leverage cross-domain relationships; for example, when recommending movies, the LLM may implicitly leverage relationships with related books and music. On the other hand, an LLM's internal knowledge is often unknown by the users who prompt the LLM (especially for proprietary models), and it might be hard to drive the model to use its implicit knowledge when re-ranking. With prompt templates T5-T8, we attempted to reduce this gap by providing the LLM with explicit item features to use during re-ranking, similarly to what is done in RAG with LLMs \cite{gao2024rag} and in the same way we do with traditional re-rankers. The results presented in the previous section show that explicit item features can sometimes benefit re-ranking. 

Greedy approaches can re-rank candidate lists composed of a high number of items (limited only by the amount of memory available), while the candidate list size when re-ranking LLM-based is limited by the size of the input context that can be fed into an LLM. Although some LLMs have long context size windows of millions of tokens (e.g.\ Google Gemini 1.5\footnote{https://deepmind.google/technologies/gemini/}) and some tricks are available to increase the size of such windows (e.g.\ \cite{ding2024longrope}), the OpenAI and Meta LLMs we used all have a limited input size of roughly 4k tokens.

Most of the traditional re-ranking approaches that we presented in Section \ref{sec:increasing_div} do not require expensive computations (they can run on CPUs) and do not have high memory requirements. Moreover, simple techniques might drastically reduce the re-ranking time (for example, item distances might be pre-computed and cached). By contrast, resource requirements are one drawback of LLMs: see the insights from the previous section. Hosting and running efficiently an LLM-based service (in our scenario, an LLM-based re-ranker) needs resources such as GPUs and several gigabytes of memory. These are expensive and might require skilled technical expertise. An LLMs' high inference time might also be a bottleneck in settings where recommendations must be generated on the fly (e.g.\ session-based recommenders). There is a growing amount of recent work on both reducing LLM inference time (e.g.\ using batch inference, flash attention, and weight sharing) and reducing their memory footprint (e.g.\ using quantization and model distillation) (see \cite{xu2023survey_compression} for a broad overview). However, the gap with traditional re-rankers remains substantial. 

When using traditional re-rankers, data and computations are typically hosted on a local infrastructure, as the cost and the technical challenges are quite low. By contrast, when running LLM-based re-ranking using a proprietary model or a proprietary platform (as we do with ChatGPT leveraging OpenAI APIs, see Section \ref{sec:experiments}), control over data and processing is generally limited. For example, data needs to be uploaded to and downloaded from the data centre where the LLM is hosted; the inference time (i.e.\ generation time) might be higher when the endpoint is overloaded; inference availability may even be disrupted; usage limits on the service might be applied; and there is need for the external platform provider to guarantee data protection. However, when using open source alternatives such as Meta's LLMs, these problems are mitigated because control over data and computations is retained.

\begin{table}[t]
\caption{Comparison of traditional re-rankers and LLM-based re-rankers beyond re-ranking performance.}
\begin{tabular}{|c|cc|}
\hline
                                       & \multicolumn{2}{c|}{\textbf{Re-rankers}}                                                \\ \hline
                                       & \multicolumn{1}{c|}{\textbf{Traditional}} & \textbf{LLM-based}                                   \\ \hline
Require explicit item features       & \multicolumn{1}{c|}{Yes}         & No (but can use them if available)                                         \\ \hline
Can re-rank large candidate lists                  & \multicolumn{1}{c|}{Yes}         & Depends on the size of the context window                  \\ \hline
Require expensive computations         & \multicolumn{1}{c|}{No}          & Yes (but can be reduced with optimizations) \\ \hline
Have a heavy memory footprint                 & \multicolumn{1}{c|}{No}          & Yes (but can be reduced with optimizations) \\ \hline
Control over data, computations, usage & \multicolumn{1}{c|}{Full}        & Limited (depending on the model )           \\ \hline
Easily address the cold-start problem  & \multicolumn{1}{c|}{Yes}         & No                                          \\ \hline
Always produce valid rankings          & \multicolumn{1}{c|}{Yes}         & No                                          \\ \hline
Easily identify items                  & \multicolumn{1}{c|}{Yes}         & No (depending on the domain)                  \\ \hline
\end{tabular}
\label{table:comparison_tradeoff}
\end{table}

From a re-ranking point of view, traditional approaches address the well-known item cold-start problem fairly smoothly: they require only that each new item be associated with its features. By contrast, LLM-based approaches might struggle with this problem. Items that are new to the system (e.g.\ newly released movies) and information related to them (such as item features or general descriptions) will be unknown to an LLM that was trained or tuned prior to release of the new item. Updating its parameters might be impractical even if fresh item information is available. Re-training will likely be periodic, possibly introducing a substantial delay before the LLM becomes aware of the new items. One solution to this problem might be the RAG approach \cite{gao2024rag}, which could be used to gather fresh contextual information related to items to enrich the re-ranking prompts. However, implementing a RAG system could be complex and expensive and would also increase the re-ranking time due to the inference time of the retrieval components.

Traditional re-rankers always produce valid rankings. By contrast, LLMs may generate text that is inconsistent, hallucinated, harmful, or irrelevant to the instructions provided in the prompt. The results of our experiments when using LLM-based re-rankers also report invalid (malformed) outputs, e.g.\ recommendations that have titles of items that do not exist, or titles that do not appear in $CL$ (see Section \ref{sec:results}). Countermeasures have been proposed in the literature to mitigate such problems \cite{ji2023survey_allucinations, zhang2023siren}, but it appears there is no one-size-fits-all solution. Advanced parsing of an LLM's output can also be implemented (and we indeed partly address invalid generations with this technique in Section \ref{sec:methodology}), but these solutions do not necessarily scale well to a wide range of invalid generation corner cases.

Traditional re-rankers typically identify items by identifiers such as product codes. This can work across different recommendation domains. In LLM-based re-ranking, the prompts we use rely on working in domains where items have useful titles to identify them. This allows the LLM to draw on its knowledge of these items. Domains such as books, movies, and music use titles in this way but this might be challenging or impractical in other domains. For example, consider domains where people are recommended to other people, such as dating or social media. An LLM may not know who the individuals are by their names. If they do know, it will be because the training data includes data about these individuals (and whether this is so is often unknown upfront, as training data is often undisclosed for state-of-the-art models). Non-uniqueness may also be a problem (the extent of the problem perhaps being greater in some domains, such as hotels, than others, such as movies). Names and titles may need to be extended (e.g.\ hotels with their cities) to reduce non-uniqueness. Finally, there are domains where there may not be titles at all, e.g.\ short posts to social media streams. When titles/names are unavailable, LLMs might still use items features or descriptions to identify items and re-rank them with proper prompts modification. In our future work we plan to explore re-ranking in this challenging scenario.

\section{Conclusion and future work}\label{sec:conclusions}

This paper investigates the use of general-purpose conversational LLMs to enhance top-$n$ recommendation diversity through re-ranking. In this paper, we proposed to prompt LLMs in a zero-shot fashion to re-rank a large set of candidate items and provide a final recommendation set. We wanted to discover whether LLMs can interpret diversification re-ranking. We also explored the impact of different prompt templates we designed: some of them aim to strike a balance between relevance and diversity, and some focus on improving diversity only. Some templates include item features as contextual information to help re-ranking, and others do not. We run experiments on two publicly available datasets (from the anime movies and books domains) where we compare LLM-based re-ranking such as ChatGPT and InstructGPT from OpenAI and Llama2-7B-Chat and Llama2-13B-Chat from Meta with traditional greedy diversification re-rankers and a random re-ranker. We analyzed their cost-performance trade-offs and provided a comprehensive discussion of the pros and cons beyond accuracy/diversity performances and computational costs to help inform the choice between LLM-based and traditional greedy approaches to perform diversification re-ranking.

First of all, our results show that the LLM-based re-rankers that we have proposed can interpret the re-ranking task. For both datasets, all re-rankers sacrifice relevance to improve diversity, but to a different extent. LLM-based rankings have better relevance/diversity trade-offs than the ones of random re-ranking but worse than the ones of the traditional re-rankers. We argue that LLM-based re-rankers are inferior to traditional re-rankers, particularly for relevance-aware metrics, because: they draw lower-ranked items from the candidate list; they are penalised by the presence of random recommendations in the final list due to invalid generation; and they sometimes struggle to effectively capture relevance and item diversity, whereas traditional re-rankers directly leverage user profiles and item features to drive re-ranking.

Re-ranking from OpenAI models are better than the ones from Meta's models: ChatGPT is, in general, the best performing LLM re-ranker, and Llama2-13B-Chat is the worst (although comparable to the others on ILD and SRecall). Our findings are in line with the majority of the literature that credits OpenAI models with more advanced capabilities than Llama models. Also, different prompt templates have different performances across LLMs and datasets, which suggests that their design should be tailored accordingly. Even so some patterns can be discerned. For example, on average, feature-aware templates benefit the re-ranking performances more than templates that do
not include item features; and templates that are designed to boost diversity rather than striking a balance between accuracy and diversity do not consistently achieve this aim.

Finally, traditional re-ranking is much faster and less resource-demanding than LLM-based re-ranking. When using LLM-based re-ranking, one should carefully consider the costs for inference and deployment and constraints on the LLM usage (such as control over data and inference rate/time).

There are many directions for future work to extend the study reported in this paper.
LLMs are improving constantly: models are getting larger, maximum context sizes are growing, and inference costs are falling. It would be valuable to repeat our study in the future. We might expect the gap between LLM-based re-ranking and traditional re-ranking to fall. With reduced costs, it would also become feasible to run experiments with more test users, a wider range of prompts, and more than two datasets.

There are also several ways to improve LLM-based re-rankers to see whether their performance can come closer to that of traditional re-rankers. One option is to train a dedicated LLM to perform the re-ranking. This can be done, for example, by combining SFT and RLHF with proper instruction/response pairs and human feedback or by deriving a distilled model \cite{gu2023knowledge_distillation}. Another option is to reduce the proportion of invalid items (and hence random recommendations) by repeating the generation or implementing more advanced inexact string matching. This might help the quality of recommendations, particularly their relevance. Improving the prompts with richer contextual information might also help. This can be done in several ways: by providing demonstrations of re-ranking in the prompts (in the spirit of few-shot learning); implementing a RAG pipeline to enrich item features; or by making the re-ranking instructions in the prompts more detailed and explicit to break down the task into individual steps (similarly to what is done in Chain-of-thoughts prompting \cite{wei2022cot}), e.g.\ \verb|Generate a new ranking where, at each step, you select an item that is different from the ones| \verb|already included in the final recommendation|.  

LLMs are essentially black-box models. 
There are only a few attempts to gain insight into what they learn  (e.g.\ \cite{anthropic_explainability}). Research into explaining how LLMs perform diversification re-ranking could be another direction for future work.

Finally, re-ranking quality could be assessed with alternative diversity metrics (e.g.\ measuring item distances on item ratings) or other beyond-accuracy metrics such as novelty and serendipity \cite{kaminskas2016diversity} to explore LLM capabilities more fully.

\begin{acks}
This publication has emanated from research supported by Science Foundation Ireland under Grant No. 12/RC/2289-P2 at the Insight SFI Research Centre for Data Analytics at UCC, which is co-funded under the European Regional Development Fund. 
\end{acks}








\bibliographystyle{ACM-Reference-Format}
\bibliography{biblio}

\appendix




% ...

% ...

\section{Examples of Prompts and outputs}

\subsection{Preliminary study}
\label{sec:appendix_prelim}
\begin{myverbatim}[title={Q1-P1. This prompt assesses whether ChatGPT can re-rank a list of anime titles into alphabetical order.}]
You are given a list of 20 anime, delimited by triple backticks.
Your task is to re-rank this list in alphabetical order.
```
0.-> Hundred
1.-> The Millionaire Detective - Balance:UNLIMITED
2.-> Freaky Highschool
3.-> Arte
4.-> Infinite Ryvius
5.-> Is this A Zombie? of the Dead
6.-> A Christmas Adventure
7.-> Junjo Romantica
8.-> Rokka:Braves of the Six Flowers
9.-> Wooser's Hand-to-Mouth Life:Awakening Arc
10.-> Jormungand Season 2:Perfect Order
11.-> Genshiken
12.-> Shirobako
13.-> Scum's Wish
14.-> Inuyashiki:Last Hero
15.-> My Name Is Teppei
16.-> Magical Project S
17.-> Hand Shakers
18.-> Tekkaman Blade
19.-> Pretty Cure
```
\end{myverbatim}

\begin{myverbatim}[title={ChatGPT's response to Q1-P1}]
```
6.-> A Christmas Adventure
3.-> Arte
2.-> Freaky Highschool
11.-> Genshiken
17.-> Hand Shakers
0.-> Hundred
14.-> Inuyashiki:Last Hero
4.-> Infinite Ryvius
5.-> Is this A Zombie? of the Dead
10.-> Jormungand Season 2:Perfect Order
7.-> Junjo Romantica
15.-> My Name Is Teppei
19.-> Pretty Cure
8.-> Rokka:Braves of the Six Flowers
13.-> Scum's Wish
12.-> Shirobako
16.-> Magical Project S
18.-> Tekkaman Blade
1.-> The Millionaire Detective - Balance:UNLIMITED
9.-> Wooser's Hand-to-Mouth Life:Awakening Arc
```
\end{myverbatim}

Only ``Magical Project S'' is out of position.

\begin{myverbatim}[title={Q1-P2. This prompt assesses whether ChatGPT can re-rank a list of anime titles by popularity.}]
You are given a list of 20 anime, delimited by triple backticks.
Your task is to re-rank the anime in this list by popularity.
```
Tsubasa RESERVoir CHRoNiCLE Season Two
El-Hazard:The Wanderers
Gintama Season 2
Beet the Vandel Buster
Umamusume:Pretty Derby
Bobobo-bo Bo-bobo
GG Bond:Ultimate Battle Prequel
Rozen Maiden:Zuruckspulen
Cardfight!! Vanguard:Shinemon Arc
Prince Mackaroo
Samurai Deeper Kyo
New Game!!
Trigun
ChaoS;Child
Beautiful Bones -Sakurako's Investigation-
The Ones Within
Tiger & Bunny
Cat Eyed Boy
Vampire Princess Miyu
Show By Rock!! 2
```
\end{myverbatim}

\begin{myverbatim}[title={ChatGPT's response to Q1-P2}]
Based on popularity, the ranked list of anime would be:

```
1. Gintama Season 2
2. Trigun
3. New Game!!
4. Tiger & Bunny
5. Tsubasa RESERVoir CHRoNiCLE Season Two
6. Umamusume:Pretty Derby
7. Rozen Maiden:Zuruckspulen
8. ChaoS;Child
9. Beautiful Bones -Sakurako's Investigation-
10. Vampire Princess Miyu
11. Samurai Deeper Kyo
12. Bobobo-bo Bo-bobo
13. Cardfight!! Vanguard:Shinemon Arc
14. The Ones Within
15. El-Hazard: The Wanderers
16. Beet the Vandel Buster
17. Show By Rock!! 2
18. GG Bond: Ultimate Battle Prequel
19. Prince Mackaroo
20. Cat Eyed Boy
```
\end{myverbatim}

We extracted the ground-truth popularity of these 20 anime movies from the dataset presented in Section \ref{sec:experiments}. The ranking proposed by ChatGPT overlaps with the ground-truth ranking by a 0.79 score according to the rank-biased-overlap metric \cite{webber2010rbo}. We ran several examples (with different item lists to re-rank), and the rank-biased-overlap scores vary between 0.5 and 0.8.

\begin{myverbatim}[title={Q1-P3. This prompt assesses whether ChatGPT can re-rank a list of anime titles by release date.}]
You are given a list of 20 anime, delimited by triple backticks.
Your task is to re-rank anime in this list by release date (from newest to
oldest).
```
Digimon Adventure 02
Chobits
Fighting Spirit
Chrono Crusade
The Great Passage
Kiba
Green Green
Poccolies
Sekko Boys
Teasing Master Takagi-san
Steel Angel Kurumi
A Centaur's Life
Persona:Trinity Soul
Magical Princess Minky Momo:Hold on to Your Dreams
Coral Investigates
The Pet Girl of Sakurasou
The Affectuous Family
Earl and Fairy
Vividred Operation
Golden Time
```
\end{myverbatim}

\begin{myverbatim}[title={ChatGPT's response to Q1-P3}]
Here is the list of anime ranked by release date from newest to oldest:

1. Poccolies
2. The Great Passage
3. Coral Investigates
4. Teasing Master Takagi-san
5. The Pet Girl of Sakurasou
6. Vividred Operation
7. Golden Time
8. A Centaur's Life
9. Persona: Trinity Soul
10. The Affectuous Family
11. Chobits
12. Steel Angel Kurumi
13. Chrono Crusade
14. Fighting Spirit
15. Kiba
16. Green Green
17. Digimon Adventure 02
18. Magical Princess Minky Momo: Hold on to Your Dreams
19. Earl and Fairy
20. Sekko Boys
\end{myverbatim}

We extracted the ground-truth item release dates from the dataset presented in Section \ref{sec:experiments}. The ranking proposed by ChatGPT overlaps with the ground-truth ranking by a 0.79 score according to the rank-biased-overlap metric. We ran several examples (with different item lists to re-rank), and the rank-biased-overlap scores vary between 0.35 and 0.7.

\begin{myverbatim}[title={Q2-P1. This prompt assesses whether ChatGPT can select the more diverse of two lists of anime titles.}]
You are given two lists of 10 anime each, list A and list B, delimited by
triple backticks.
Which of the two lists is more diverse?

```
LIST A:
Mr. Stain on Junk Alley
Let's Dance with Papa
King of Classroom, Yamazaki
Poo-Neko
The Adventures of the Little Koala
Oscar's Oasis
The Affectuous Family
Little Wansa
Hippo and Thomas
Weekly Shimako

LIST B:
Hundred
The Millionaire Detective - Balance:UNLIMITED
Freaky Highschool
Arte
Infinite Ryvius
Is this A Zombie? of the Dead
A Christmas Adventure
Junjo Romantica
Rokka:Braves of the Six Flowers
Wooser's Hand-to-Mouth Life:Awakening Arc
Jormungand Season 2:Perfect Order
```
\end{myverbatim}

\begin{myverbatim}[title={ChatGPT's response to Q2-P1}]
Diversity in anime can be assessed based on various factors such as genre,
themes, art style, target audience, and more.
Let's analyze the two lists in terms of genre diversity since it's a common way
to measure variety in anime.

List A:

    Mr. Stain on Junk Alley: Stop-motion animation, comedy
    Let's Dance with Papa: Slice of life, comedy
    King of Classroom, Yamazaki: Comedy, school
    Poo-Neko: Adventure, fantasy
    The Adventures of the Little Koala: Adventure, children
    Oscar's Oasis: Adventure, comedy
    The Affectionate Family: Comedy, family
    Little Wansa: Adventure, children
    Hippo and Thomas: Adventure, children
    Weekly Shimako: Slice of life, comedy

List B:

    Hundred: Action, harem
    The Millionaire Detective - Balance:UNLIMITED: Mystery, drama
    Freaky Highschool: Supernatural, comedy
    Arte: Historical, drama
    Infinite Ryvius: Sci-fi, drama
    Is this A Zombie? of the Dead: Action, comedy, supernatural
    A Christmas Adventure: Christmas, adventure
    Junjo Romantica: Romance, drama
    Rokka: Braves of the Six Flowers: Fantasy, mystery
    Wooser's Hand-to-Mouth Life: Awakening Arc: Comedy, slice of life
    Jormungand Season 2: Perfect Order: Action, adventure

In terms of genre diversity, both lists showcase a variety of genres.
However, List B appears to have a broader range of genres, including
action, mystery, supernatural, historical, romance, and Christmas-themed
anime. List A, on the other hand, leans more towards adventure, comedy,
and slice of life genres. Therefore, List B is more diverse in terms of genres.
\end{myverbatim}

The answer is ``correct'' and consistent with a genre-based diversity perspective because the set of genres that ChatGPT associates with the movies in list B is larger than the set of genres it associates with the movies in list A. This is also ``correct'' if we consider the ground-truth (the genres in the Anime dataset): in the dataset, the movies in list B are assigned a handful of genres, whereas the movies in list A all have one genre only (``comedy'').

\begin{myverbatim}[title={Q2-P2. This prompt assesses whether ChatGPT can select the more diverse of two lists of anime titles.}]
You are given two lists of 10 anime each, list A and list B, delimited by
triple backticks.
Which of the two lists is more diverse?

```
LIST A:
Ladies versus Butlers!
La storia della Arcana Famiglia
Sirius the Jaeger
School Babysitters
Pop Team Epic
Servant x Service
Uta no Prince Sama
Dragonar Academy
NAKAIMO - My Little Sister Is Among Them!
His and Her Circumstances

LIST B:
Jormungand Season 2:Perfect Order
Genshiken
Shirobako
Scum's Wish
Inuyashiki:Last Hero
My Name Is Teppei
Magical Project S
Hand Shakers
Tekkaman Blade
Pretty Cure
```
\end{myverbatim}

\begin{myverbatim}[title={ChatGPT's response to Q2-P2}]
To determine which list is more diverse, we can consider the variety of genres, 
themes, and settings present in each list. It's important to note that 
diversity can be subjective, and different criteria can be used to evaluate it. 
Here, I'll consider the diversity based on the variety of genres and themes.

List A:
1. Ladies versus Butlers! - Ecchi, Comedy, Romance
2. La storia della Arcana Famiglia - Action, Mystery, Romance, Supernatural
3. Sirius the Jaeger - Action, Supernatural, Historical, Vampire
4. School Babysitters - Slice of Life, Comedy
5. Pop Team Epic - Comedy, Parody
6. Servant x Service - Slice of Life, Comedy
7. Uta no Prince Sama - Music, Romance
8. Dragonar Academy - Action, Fantasy, Romance
9. NAKAIMO - My Little Sister Is Among Them! - Harem, Comedy, Romance
10. His and Her Circumstances - Romance, Drama, Comedy, School

List B:
1. Jormungand Season 2: Perfect Order - Action, Adventure
2. Genshiken - Slice of Life, Comedy
3. Shirobako - Slice of Life, Comedy, Drama
4. Scum's Wish - Drama, Romance
5. Inuyashiki: Last Hero - Sci-Fi, Action, Drama
6. My Name Is Teppei - Comedy, Sports
7. Magical Project S - Magic, Comedy, Fantasy
8. Hand Shakers - Action, Supernatural
9. Tekkaman Blade - Action, Adventure, Mecha
10. Pretty Cure - Mahou Shoujo, Magic, Fantasy

Observations:
- List A covers a variety of genres, including ecchi, action, slice of life, 
comedy, romance, and supernatural.
- List B also offers a mix of genres, including action, slice of life, comedy, 
drama, sci-fi, sports, magic, and mecha.

Both lists display diversity, but List A may have a slightly broader range of 
genres. It includes ecchi, which is not represented in List B. However, 
diversity can be subjective, and different viewers may have different
preferences for what they consider diverse.
\end{myverbatim}

Again, the answer is ``correct'', and for the same reasons as for Q2-P2.

\begin{myverbatim}[title={Q2-P3. This prompt assesses whether ChatGPT can select the more diverse of two lists of anime titles.}]
You are given two lists of 10 anime each, list A and list B, delimited by
triple backticks.
Which of the two lists is more diverse?

```
LIST A:

The Millionaire Detective - Balance:UNLIMITED
Freaky Highschool
Arte
Infinite Ryvius
Is this A Zombie? of the Dead
A Christmas Adventure
Junjo Romantica
Rokka:Braves of the Six Flowers
Wooser's Hand-to-Mouth Life:Awakening Arc

LIST B:
Jormungand Season 2:Perfect Order
Genshiken
Shirobako
Scum's Wish
Inuyashiki:Last Hero
My Name Is Teppei
Magical Project S
Hand Shakers
Tekkaman Blade
Pretty Cure
```
\end{myverbatim}

\begin{myverbatim}[title={ChatGPT's response to Q2-P3}]
To assess diversity in anime, we can consider various factors such as genre,
themes, art style, and target audience.
Let's analyze the two lists in terms of genre diversity:

List A:

    The Millionaire Detective - Balance:UNLIMITED: Mystery, drama
    Freaky Highschool: Supernatural, comedy
    Arte: Historical, drama
    Infinite Ryvius: Sci-fi, drama
    Is this A Zombie? of the Dead: Action, comedy, supernatural
    A Christmas Adventure: Christmas, adventure
    Junjo Romantica: Romance, drama
    Rokka: Braves of the Six Flowers: Fantasy, mystery
    Wooser's Hand-to-Mouth Life: Awakening Arc: Comedy, slice of life

List B:

    Jormungand Season 2: Perfect Order: Action, adventure
    Genshiken: Comedy, slice of life
    Shirobako: Slice of life, comedy
    Scum's Wish: Drama, romance
    Inuyashiki: Last Hero: Action, drama, sci-fi
    My Name Is Teppei: Sports, drama
    Magical Project S: Magic, comedy
    Hand Shakers: Action, drama, fantasy
    Tekkaman Blade: Action, sci-fi
    Pretty Cure: Action, adventure, magic

In terms of genre diversity, both lists showcase a variety of genres. List B,
however, seems to have a broader range, including action, adventure,
comedy, slice of life, drama, romance, sci-fi, magic, and sports. List A
also covers multiple genres, but it has a stronger emphasis on drama and
romance, with fewer distinct genres represented. Therefore, List B is more
diverse in terms of genres.
\end{myverbatim}

This time the answer is ``incorrect'' as list A is more diverse than list B, according both to the genres that ChatGPT associates with the two lists and the genres extracted from the Anime dataset. 


\subsection{Diversification re-ranking}\label{sec:appendix_rerank}

\begin{myverbatim}[title={Example of a prompt from template T1 and using the anime dataset.}]
You are given a ranked recommendation list of 40 items for a user, 
delimited by triple backticks.
Your task is to re-rank this candidate list and provide a final top-10 
recommendation list where the goal is to balance relevance and diversity. 
Strictly use the following format for the output, and don't provide additional 
information.

1-> <anime name>
2-> <anime name>
...
10-> <anime name>

```
1. Dragon Ball Z
2. Naruto
3. Deadman Wonderland
...
9. Code Geass:Lelouch of the Rebellion R2
10. One Punch Man
...
15. Haikyu!!
...
22. Made in Abyss
...
27. Death Parade
...
30. Attack on Titan Season 3 Part 2
...
32. Re:ZERO -Starting Life in Another World
...
34. The Promised Neverland
35. Demon Slayer:Kimetsu no Yaiba
...
40. Your Lie in April
```
\end{myverbatim}

\begin{myverbatim}[title={ChatGPT's output for the previous prompt.}]
1-> Deadman Wonderland
2-> Code Geass:Lelouch of the Rebellion R2
3-> One Punch Man
4-> Death Parade
5-> Attack on Titan Season 3 Part 2
6-> Re:ZERO -Starting Life in Another World
7-> Made in Abyss
8-> Haikyu!!
9-> The Promised Neverland
10-> Demon Slayer:Kimetsu no Yaiba
\end{myverbatim}

\section{Results}

\begin{table}[h]
\small
\caption{Re-ranking performance for the Anime dataset. All metrics are computed with a cutoff of size $n =|RL|=10$. For all metrics, the higher the scores, the better the performance. For each measure, we also report the confidence intervals calculated with a 95\% confidence level.}
\begin{tabular}{cccccccc}
\cline{3-8}
\multicolumn{1}{l}{}                               & \multicolumn{1}{c|}{}                  & \multicolumn{1}{c|}{\textbf{NDCG}}       & \multicolumn{1}{c|}{\textbf{$\alpha$-NDCG}}    & \multicolumn{1}{c|}{\textbf{EILD}}       & \multicolumn{1}{c|}{\textbf{ILD}}        & \multicolumn{1}{c|}{\textbf{rSRecall}}   & \multicolumn{1}{c|}{\textbf{SRecall}}    \\ \hline
\multicolumn{1}{|c|}{MF}                           & \multicolumn{1}{c|}{}                  & \multicolumn{1}{c|}{0.316  $\pm$  0.020} & \multicolumn{1}{c|}{0.325  $\pm$  0.019} & \multicolumn{1}{c|}{0.221  $\pm$  0.015} & \multicolumn{1}{c|}{0.776  $\pm$  0.006} & \multicolumn{1}{c|}{0.234  $\pm$  0.012} & \multicolumn{1}{c|}{0.473  $\pm$  0.006} \\ \hline
\multicolumn{1}{|c|}{\textbf{Re-ranker}}           & \multicolumn{1}{c|}{\textbf{Template}} & \multicolumn{1}{c|}{}                    & \multicolumn{1}{c|}{}                    & \multicolumn{1}{c|}{}                    & \multicolumn{1}{c|}{}                    & \multicolumn{1}{c|}{}                    & \multicolumn{1}{c|}{}                    \\ \hline
\multicolumn{1}{|c|}{Random}                       & \multicolumn{1}{c|}{}                  & \multicolumn{1}{c|}{0.130  $\pm$  0.012} & \multicolumn{1}{c|}{0.154  $\pm$  0.013} & \multicolumn{1}{c|}{0.094  $\pm$  0.011} & \multicolumn{1}{c|}{0.811  $\pm$  0.004} & \multicolumn{1}{c|}{0.145  $\pm$  0.010} & \multicolumn{1}{c|}{0.499  $\pm$  0.005} \\ \hline
\multicolumn{1}{|c|}{MMR}                          & \multicolumn{1}{c|}{}                  & \multicolumn{1}{c|}{0.299  $\pm$  0.019} & \multicolumn{1}{c|}{0.313  $\pm$  0.018} & \multicolumn{1}{c|}{0.222  $\pm$  0.015} & \multicolumn{1}{c|}{0.874  $\pm$  0.003} & \multicolumn{1}{c|}{0.237  $\pm$  0.013} & \multicolumn{1}{c|}{0.544  $\pm$  0.005} \\ \hline
\multicolumn{1}{|c|}{xQuad}                        & \multicolumn{1}{c|}{}                  & \multicolumn{1}{c|}{0.301  $\pm$  0.019} & \multicolumn{1}{c|}{0.350  $\pm$  0.021} & \multicolumn{1}{c|}{0.212  $\pm$  0.015} & \multicolumn{1}{c|}{0.780  $\pm$  0.005} & \multicolumn{1}{c|}{0.251  $\pm$  0.013} & \multicolumn{1}{c|}{0.539  $\pm$  0.006} \\ \hline
\multicolumn{1}{|c|}{RxQuad}                       & \multicolumn{1}{c|}{}                  & \multicolumn{1}{c|}{0.295  $\pm$  0.019} & \multicolumn{1}{c|}{0.338  $\pm$  0.020} & \multicolumn{1}{c|}{0.207  $\pm$  0.014} & \multicolumn{1}{c|}{0.770  $\pm$  0.005} & \multicolumn{1}{c|}{0.244  $\pm$  0.013} & \multicolumn{1}{c|}{0.506  $\pm$  0.006} \\ \hline
                                                   & \textbf{}                              &                                          &                                          &                                          &                                          &                                          &                                          \\ \hline
\multicolumn{1}{|c|}{\multirow{8}{*}{ChatGPT}}     & \multicolumn{1}{c|}{T1}                & \multicolumn{1}{c|}{0.279  $\pm$  0.018} & \multicolumn{1}{c|}{0.298  $\pm$  0.018} & \multicolumn{1}{c|}{0.192  $\pm$  0.014} & \multicolumn{1}{c|}{0.812  $\pm$  0.005} & \multicolumn{1}{c|}{0.223  $\pm$  0.012} & \multicolumn{1}{c|}{0.509  $\pm$  0.005} \\ \cline{2-8} 
\multicolumn{1}{|c|}{}                             & \multicolumn{1}{c|}{T2}                & \multicolumn{1}{c|}{0.298  $\pm$  0.019} & \multicolumn{1}{c|}{0.314  $\pm$  0.018} & \multicolumn{1}{c|}{0.213  $\pm$  0.015} & \multicolumn{1}{c|}{0.812  $\pm$  0.005} & \multicolumn{1}{c|}{0.232  $\pm$  0.012} & \multicolumn{1}{c|}{0.503  $\pm$  0.006} \\ \cline{2-8} 
\multicolumn{1}{|c|}{}                             & \multicolumn{1}{c|}{T3}                & \multicolumn{1}{c|}{0.282  $\pm$  0.018} & \multicolumn{1}{c|}{0.305  $\pm$  0.018} & \multicolumn{1}{c|}{0.198  $\pm$  0.014} & \multicolumn{1}{c|}{0.817  $\pm$  0.005} & \multicolumn{1}{c|}{0.227  $\pm$  0.012} & \multicolumn{1}{c|}{0.513  $\pm$  0.005} \\ \cline{2-8} 
\multicolumn{1}{|c|}{}                             & \multicolumn{1}{c|}{T4}                & \multicolumn{1}{c|}{0.260  $\pm$  0.017} & \multicolumn{1}{c|}{0.286  $\pm$  0.018} & \multicolumn{1}{c|}{0.179  $\pm$  0.014} & \multicolumn{1}{c|}{0.815  $\pm$  0.004} & \multicolumn{1}{c|}{0.217  $\pm$  0.012} & \multicolumn{1}{c|}{0.519  $\pm$  0.005} \\ \cline{2-8} 
\multicolumn{1}{|c|}{}                             & \multicolumn{1}{c|}{T5}                & \multicolumn{1}{c|}{0.284  $\pm$  0.019} & \multicolumn{1}{c|}{0.306  $\pm$  0.018} & \multicolumn{1}{c|}{0.201  $\pm$  0.015} & \multicolumn{1}{c|}{0.812  $\pm$  0.004} & \multicolumn{1}{c|}{0.228  $\pm$  0.013} & \multicolumn{1}{c|}{0.510  $\pm$  0.005} \\ \cline{2-8} 
\multicolumn{1}{|c|}{}                             & \multicolumn{1}{c|}{T6}                & \multicolumn{1}{c|}{0.278  $\pm$  0.018} & \multicolumn{1}{c|}{0.303  $\pm$  0.018} & \multicolumn{1}{c|}{0.197  $\pm$  0.014} & \multicolumn{1}{c|}{0.816  $\pm$  0.004} & \multicolumn{1}{c|}{0.228  $\pm$  0.012} & \multicolumn{1}{c|}{0.518  $\pm$  0.005} \\ \cline{2-8} 
\multicolumn{1}{|c|}{}                             & \multicolumn{1}{c|}{T7}                & \multicolumn{1}{c|}{0.284  $\pm$  0.018} & \multicolumn{1}{c|}{0.307  $\pm$  0.018} & \multicolumn{1}{c|}{0.202  $\pm$  0.015} & \multicolumn{1}{c|}{0.807  $\pm$  0.005} & \multicolumn{1}{c|}{0.231  $\pm$  0.012} & \multicolumn{1}{c|}{0.507  $\pm$  0.005} \\ \cline{2-8} 
\multicolumn{1}{|c|}{}                             & \multicolumn{1}{c|}{T8}                & \multicolumn{1}{c|}{0.307  $\pm$  0.019} & \multicolumn{1}{c|}{0.322  $\pm$  0.019} & \multicolumn{1}{c|}{0.219  $\pm$  0.015} & \multicolumn{1}{c|}{0.806  $\pm$  0.005} & \multicolumn{1}{c|}{0.237  $\pm$  0.012} & \multicolumn{1}{c|}{0.500  $\pm$  0.005} \\ \hline
\multicolumn{1}{l}{}                               & \multicolumn{1}{l}{}                   & \multicolumn{1}{l}{}                     & \multicolumn{1}{l}{}                     & \multicolumn{1}{l}{}                     & \multicolumn{1}{l}{}                     & \multicolumn{1}{l}{}                     & \multicolumn{1}{l}{}                     \\ \hline
\multicolumn{1}{|c|}{\multirow{8}{*}{InstructGPT}} & \multicolumn{1}{c|}{T1}                & \multicolumn{1}{c|}{0.265  $\pm$  0.018} & \multicolumn{1}{c|}{0.289  $\pm$  0.018} & \multicolumn{1}{c|}{0.187  $\pm$  0.014} & \multicolumn{1}{c|}{0.812  $\pm$  0.005} & \multicolumn{1}{c|}{0.221  $\pm$  0.012} & \multicolumn{1}{c|}{0.511  $\pm$  0.005} \\ \cline{2-8} 
\multicolumn{1}{|c|}{}                             & \multicolumn{1}{c|}{T2}                & \multicolumn{1}{c|}{0.276  $\pm$  0.018} & \multicolumn{1}{c|}{0.295  $\pm$  0.018} & \multicolumn{1}{c|}{0.189  $\pm$  0.014} & \multicolumn{1}{c|}{0.813  $\pm$  0.005} & \multicolumn{1}{c|}{0.219  $\pm$  0.012} & \multicolumn{1}{c|}{0.506  $\pm$  0.006} \\ \cline{2-8} 
\multicolumn{1}{|c|}{}                             & \multicolumn{1}{c|}{T3}                & \multicolumn{1}{c|}{0.263  $\pm$  0.017} & \multicolumn{1}{c|}{0.286  $\pm$  0.018} & \multicolumn{1}{c|}{0.183  $\pm$  0.014} & \multicolumn{1}{c|}{0.818  $\pm$  0.005} & \multicolumn{1}{c|}{0.216  $\pm$  0.012} & \multicolumn{1}{c|}{0.515  $\pm$  0.005} \\ \cline{2-8} 
\multicolumn{1}{|c|}{}                             & \multicolumn{1}{c|}{T4}                & \multicolumn{1}{c|}{0.261  $\pm$  0.017} & \multicolumn{1}{c|}{0.286  $\pm$  0.018} & \multicolumn{1}{c|}{0.184  $\pm$  0.014} & \multicolumn{1}{c|}{0.814  $\pm$  0.004} & \multicolumn{1}{c|}{0.218  $\pm$  0.012} & \multicolumn{1}{c|}{0.512  $\pm$  0.005} \\ \cline{2-8} 
\multicolumn{1}{|c|}{}                             & \multicolumn{1}{c|}{T5}                & \multicolumn{1}{c|}{0.274  $\pm$  0.018} & \multicolumn{1}{c|}{0.298  $\pm$  0.018} & \multicolumn{1}{c|}{0.193  $\pm$  0.014} & \multicolumn{1}{c|}{0.814  $\pm$  0.005} & \multicolumn{1}{c|}{0.223  $\pm$  0.012} & \multicolumn{1}{c|}{0.507  $\pm$  0.005} \\ \cline{2-8} 
\multicolumn{1}{|c|}{}                             & \multicolumn{1}{c|}{T6}                & \multicolumn{1}{c|}{0.278  $\pm$  0.018} & \multicolumn{1}{c|}{0.300  $\pm$  0.018} & \multicolumn{1}{c|}{0.194  $\pm$  0.014} & \multicolumn{1}{c|}{0.813  $\pm$  0.004} & \multicolumn{1}{c|}{0.223  $\pm$  0.012} & \multicolumn{1}{c|}{0.507  $\pm$  0.005} \\ \cline{2-8} 
\multicolumn{1}{|c|}{}                             & \multicolumn{1}{c|}{T7}                & \multicolumn{1}{c|}{0.270  $\pm$  0.018} & \multicolumn{1}{c|}{0.293  $\pm$  0.018} & \multicolumn{1}{c|}{0.189  $\pm$  0.014} & \multicolumn{1}{c|}{0.800  $\pm$  0.005} & \multicolumn{1}{c|}{0.219  $\pm$  0.012} & \multicolumn{1}{c|}{0.496  $\pm$  0.006} \\ \cline{2-8} 
\multicolumn{1}{|c|}{}                             & \multicolumn{1}{c|}{T8}                & \multicolumn{1}{c|}{0.264  $\pm$  0.018} & \multicolumn{1}{c|}{0.284  $\pm$  0.018} & \multicolumn{1}{c|}{0.189  $\pm$  0.014} & \multicolumn{1}{c|}{0.813  $\pm$  0.004} & \multicolumn{1}{c|}{0.216  $\pm$  0.012} & \multicolumn{1}{c|}{0.503  $\pm$  0.005} \\ \hline
\multicolumn{1}{l}{}                               & \multicolumn{1}{l}{}                   & \multicolumn{1}{l}{}                     &                                          &                                          &                                          &                                          &                                          \\ \hline
\multicolumn{1}{|c|}{\multirow{8}{*}{Llama2-7B}}   & \multicolumn{1}{c|}{T1}                & \multicolumn{1}{l|}{0.252  $\pm$  0.018} & \multicolumn{1}{l|}{0.280  $\pm$  0.018} & \multicolumn{1}{l|}{0.172  $\pm$  0.014} & \multicolumn{1}{l|}{0.814  $\pm$  0.004} & \multicolumn{1}{l|}{0.212  $\pm$  0.012} & \multicolumn{1}{c|}{0.515  $\pm$  0.005} \\ \cline{2-8} 
\multicolumn{1}{|c|}{}                             & \multicolumn{1}{c|}{T2}                & \multicolumn{1}{l|}{0.236  $\pm$  0.017} & \multicolumn{1}{l|}{0.260  $\pm$  0.017} & \multicolumn{1}{l|}{0.158  $\pm$  0.013} & \multicolumn{1}{l|}{0.819  $\pm$  0.004} & \multicolumn{1}{l|}{0.198  $\pm$  0.012} & \multicolumn{1}{c|}{0.506  $\pm$  0.005} \\ \cline{2-8} 
\multicolumn{1}{|c|}{}                             & \multicolumn{1}{c|}{T3}                & \multicolumn{1}{l|}{0.244  $\pm$  0.017} & \multicolumn{1}{l|}{0.272  $\pm$  0.018} & \multicolumn{1}{l|}{0.160  $\pm$  0.014} & \multicolumn{1}{l|}{0.818  $\pm$  0.004} & \multicolumn{1}{l|}{0.202  $\pm$  0.012} & \multicolumn{1}{c|}{0.512  $\pm$  0.005} \\ \cline{2-8} 
\multicolumn{1}{|c|}{}                             & \multicolumn{1}{c|}{T4}                & \multicolumn{1}{l|}{0.254  $\pm$  0.017} & \multicolumn{1}{l|}{0.283  $\pm$  0.018} & \multicolumn{1}{c|}{0.173  $\pm$  0.014} & \multicolumn{1}{c|}{0.812  $\pm$  0.004} & \multicolumn{1}{c|}{0.216  $\pm$  0.012} & \multicolumn{1}{c|}{0.516  $\pm$  0.005} \\ \cline{2-8} 
\multicolumn{1}{|c|}{}                             & \multicolumn{1}{c|}{T5}                & \multicolumn{1}{l|}{0.236  $\pm$  0.017} & \multicolumn{1}{l|}{0.269  $\pm$  0.018} & \multicolumn{1}{c|}{0.163  $\pm$  0.014} & \multicolumn{1}{c|}{0.806  $\pm$  0.005} & \multicolumn{1}{c|}{0.207  $\pm$  0.012} & \multicolumn{1}{c|}{0.513  $\pm$  0.005} \\ \cline{2-8} 
\multicolumn{1}{|c|}{}                             & \multicolumn{1}{c|}{T6}                & \multicolumn{1}{l|}{0.243  $\pm$  0.017} & \multicolumn{1}{c|}{0.272  $\pm$  0.018} & \multicolumn{1}{c|}{0.169  $\pm$  0.014} & \multicolumn{1}{c|}{0.806  $\pm$  0.005} & \multicolumn{1}{c|}{0.207  $\pm$  0.012} & \multicolumn{1}{c|}{0.512  $\pm$  0.005} \\ \cline{2-8} 
\multicolumn{1}{|c|}{}                             & \multicolumn{1}{c|}{T7}                & \multicolumn{1}{l|}{0.262  $\pm$  0.017} & \multicolumn{1}{c|}{0.283  $\pm$  0.018} & \multicolumn{1}{c|}{0.183  $\pm$  0.014} & \multicolumn{1}{c|}{0.809  $\pm$  0.005} & \multicolumn{1}{c|}{0.217  $\pm$  0.012} & \multicolumn{1}{c|}{0.506  $\pm$  0.006} \\ \cline{2-8} 
\multicolumn{1}{|c|}{}                             & \multicolumn{1}{c|}{T8}                & \multicolumn{1}{l|}{0.256  $\pm$  0.018} & \multicolumn{1}{c|}{0.274  $\pm$  0.017} & \multicolumn{1}{c|}{0.179  $\pm$  0.014} & \multicolumn{1}{c|}{0.812  $\pm$  0.005} & \multicolumn{1}{c|}{0.208  $\pm$  0.012} & \multicolumn{1}{c|}{0.505  $\pm$  0.006} \\ \hline
                                                   &                                        & \multicolumn{1}{l}{}                     &                                          &                                          &                                          &                                          &                                          \\ \hline
\multicolumn{1}{|l|}{\multirow{8}{*}{Llama2-13B}}  & \multicolumn{1}{c|}{T1}                & \multicolumn{1}{l|}{0.255  $\pm$  0.018} & \multicolumn{1}{c|}{0.276  $\pm$  0.018} & \multicolumn{1}{c|}{0.180  $\pm$  0.014} & \multicolumn{1}{c|}{0.810  $\pm$  0.004} & \multicolumn{1}{c|}{0.213  $\pm$  0.012} & \multicolumn{1}{c|}{0.499  $\pm$  0.006} \\ \cline{2-8} 
\multicolumn{1}{|l|}{}                             & \multicolumn{1}{c|}{T2}                & \multicolumn{1}{l|}{0.168  $\pm$  0.015} & \multicolumn{1}{c|}{0.186  $\pm$  0.015} & \multicolumn{1}{c|}{0.120  $\pm$  0.012} & \multicolumn{1}{c|}{0.818  $\pm$  0.004} & \multicolumn{1}{c|}{0.160  $\pm$  0.011} & \multicolumn{1}{c|}{0.487  $\pm$  0.005} \\ \cline{2-8} 
\multicolumn{1}{|l|}{}                             & \multicolumn{1}{c|}{T3}                & \multicolumn{1}{l|}{0.226  $\pm$  0.017} & \multicolumn{1}{c|}{0.244  $\pm$  0.017} & \multicolumn{1}{c|}{0.163  $\pm$  0.013} & \multicolumn{1}{c|}{0.815  $\pm$  0.005} & \multicolumn{1}{c|}{0.196  $\pm$  0.012} & \multicolumn{1}{c|}{0.497  $\pm$  0.006} \\ \cline{2-8} 
\multicolumn{1}{|l|}{}                             & \multicolumn{1}{c|}{T4}                & \multicolumn{1}{l|}{0.257  $\pm$  0.017} & \multicolumn{1}{c|}{0.279  $\pm$  0.018} & \multicolumn{1}{c|}{0.182  $\pm$  0.014} & \multicolumn{1}{c|}{0.812  $\pm$  0.005} & \multicolumn{1}{c|}{0.216  $\pm$  0.012} & \multicolumn{1}{c|}{0.502  $\pm$  0.006} \\ \cline{2-8} 
\multicolumn{1}{|l|}{}                             & \multicolumn{1}{c|}{T5}                & \multicolumn{1}{l|}{0.257  $\pm$  0.018} & \multicolumn{1}{c|}{0.280  $\pm$  0.018} & \multicolumn{1}{c|}{0.179  $\pm$  0.014} & \multicolumn{1}{c|}{0.808  $\pm$  0.005} & \multicolumn{1}{c|}{0.212  $\pm$  0.012} & \multicolumn{1}{c|}{0.497  $\pm$  0.006} \\ \cline{2-8} 
\multicolumn{1}{|l|}{}                             & \multicolumn{1}{c|}{T6}                & \multicolumn{1}{l|}{0.247  $\pm$  0.017} & \multicolumn{1}{c|}{0.267  $\pm$  0.018} & \multicolumn{1}{c|}{0.171  $\pm$  0.014} & \multicolumn{1}{c|}{0.809  $\pm$  0.005} & \multicolumn{1}{c|}{0.204  $\pm$  0.012} & \multicolumn{1}{c|}{0.497  $\pm$  0.006} \\ \cline{2-8} 
\multicolumn{1}{|l|}{}                             & \multicolumn{1}{c|}{T7}                & \multicolumn{1}{l|}{0.266  $\pm$  0.018} & \multicolumn{1}{l|}{0.288  $\pm$  0.018} & \multicolumn{1}{l|}{0.185  $\pm$  0.014} & \multicolumn{1}{l|}{0.801  $\pm$  0.005} & \multicolumn{1}{c|}{0.215  $\pm$  0.012} & \multicolumn{1}{c|}{0.492  $\pm$  0.006} \\ \cline{2-8} 
\multicolumn{1}{|l|}{}                             & \multicolumn{1}{c|}{T8}                & \multicolumn{1}{l|}{0.195  $\pm$  0.015} & \multicolumn{1}{l|}{0.217  $\pm$  0.016} & \multicolumn{1}{l|}{0.145  $\pm$  0.013} & \multicolumn{1}{l|}{0.808  $\pm$  0.005} & \multicolumn{1}{c|}{0.180  $\pm$  0.011} & \multicolumn{1}{c|}{0.489  $\pm$  0.005} \\ \hline
\end{tabular}
\label{table:anime_full_results}
\end{table}

\begin{table}[h]
\small
\caption{Re-ranking performance for the Goodreads dataset. All metrics are computed with a cutoff of size $n =|RL|=10$. For all metrics, the higher the scores, the better the performance. For each measure, we also report the confidence intervals calculated with a 95\% confidence level.}
\begin{tabular}{ccllllll}
\cline{3-8}
\multicolumn{1}{l}{}                               & \multicolumn{1}{l|}{}                  & \multicolumn{1}{c|}{\textbf{NDCG}}       & \multicolumn{1}{c|}{\textbf{$\alpha$-NDCG}}    & \multicolumn{1}{c|}{\textbf{EILD}}       & \multicolumn{1}{c|}{\textbf{ILD}}        & \multicolumn{1}{c|}{\textbf{rSRecall}}   & \multicolumn{1}{c|}{\textbf{SRecall}}    \\ \hline
\multicolumn{1}{|c|}{MF}                           & \multicolumn{1}{c|}{}                  & \multicolumn{1}{c|}{0.249  $\pm$  0.022} & \multicolumn{1}{c|}{0.271  $\pm$  0.022} & \multicolumn{1}{c|}{0.092  $\pm$  0.010} & \multicolumn{1}{c|}{0.689  $\pm$  0.006} & \multicolumn{1}{c|}{0.355  $\pm$  0.025} & \multicolumn{1}{c|}{0.871  $\pm$  0.009} \\ \hline
\multicolumn{1}{|c|}{\textbf{Re-ranker}}           & \multicolumn{1}{c|}{\textbf{Template}} & \multicolumn{1}{c|}{}                    & \multicolumn{1}{l|}{}                    & \multicolumn{1}{l|}{}                    & \multicolumn{1}{l|}{}                    & \multicolumn{1}{l|}{}                    & \multicolumn{1}{l|}{}                    \\ \hline
\multicolumn{1}{|c|}{Random}                       & \multicolumn{1}{c|}{}                  & \multicolumn{1}{c|}{0.084  $\pm$  0.010} & \multicolumn{1}{l|}{0.109  $\pm$  0.013} & \multicolumn{1}{l|}{0.026  $\pm$  0.006} & \multicolumn{1}{l|}{0.702  $\pm$  0.006} & \multicolumn{1}{l|}{0.199  $\pm$  0.021} & \multicolumn{1}{l|}{0.889  $\pm$  0.009} \\ \hline
\multicolumn{1}{|c|}{MMR}                          & \multicolumn{1}{c|}{}                  & \multicolumn{1}{c|}{0.237  $\pm$  0.020} & \multicolumn{1}{l|}{0.253  $\pm$  0.021} & \multicolumn{1}{l|}{0.090  $\pm$  0.011} & \multicolumn{1}{l|}{0.810  $\pm$  0.004} & \multicolumn{1}{l|}{0.336  $\pm$  0.025} & \multicolumn{1}{l|}{0.934  $\pm$  0.006} \\ \hline
\multicolumn{1}{|c|}{xQuad}                        & \multicolumn{1}{c|}{}                  & \multicolumn{1}{c|}{0.229  $\pm$  0.019} & \multicolumn{1}{l|}{0.308  $\pm$  0.024} & \multicolumn{1}{l|}{0.084  $\pm$  0.009} & \multicolumn{1}{l|}{0.632  $\pm$  0.006} & \multicolumn{1}{l|}{0.386  $\pm$  0.026} & \multicolumn{1}{l|}{0.958  $\pm$  0.005} \\ \hline
\multicolumn{1}{|c|}{RxQuad}                       & \multicolumn{1}{c|}{}                  & \multicolumn{1}{c|}{0.229  $\pm$  0.019} & \multicolumn{1}{l|}{0.303  $\pm$  0.024} & \multicolumn{1}{l|}{0.084  $\pm$  0.009} & \multicolumn{1}{l|}{0.635  $\pm$  0.006} & \multicolumn{1}{l|}{0.381  $\pm$  0.026} & \multicolumn{1}{l|}{0.946  $\pm$  0.006} \\ \hline
                                                   &                                        &                                          &                                          &                                          &                                          &                                          &                                          \\ \hline
\multicolumn{1}{|c|}{\multirow{8}{*}{ChatGPT}}     & \multicolumn{1}{c|}{T1}                & \multicolumn{1}{l|}{0.161  $\pm$  0.018} & \multicolumn{1}{l|}{0.210  $\pm$  0.022} & \multicolumn{1}{l|}{0.045  $\pm$  0.008} & \multicolumn{1}{l|}{0.655  $\pm$  0.007} & \multicolumn{1}{l|}{0.269  $\pm$  0.025} & \multicolumn{1}{l|}{0.877  $\pm$  0.009} \\ \cline{2-8} 
\multicolumn{1}{|c|}{}                             & \multicolumn{1}{c|}{T2}                & \multicolumn{1}{l|}{0.180  $\pm$  0.018} & \multicolumn{1}{l|}{0.227  $\pm$  0.022} & \multicolumn{1}{l|}{0.059  $\pm$  0.009} & \multicolumn{1}{l|}{0.672  $\pm$  0.007} & \multicolumn{1}{l|}{0.296  $\pm$  0.025} & \multicolumn{1}{l|}{0.886  $\pm$  0.008} \\ \cline{2-8} 
\multicolumn{1}{|c|}{}                             & \multicolumn{1}{c|}{T3}                & \multicolumn{1}{l|}{0.156  $\pm$  0.017} & \multicolumn{1}{l|}{0.209  $\pm$  0.022} & \multicolumn{1}{l|}{0.046  $\pm$  0.007} & \multicolumn{1}{l|}{0.652  $\pm$  0.007} & \multicolumn{1}{l|}{0.266  $\pm$  0.024} & \multicolumn{1}{l|}{0.883  $\pm$  0.009} \\ \cline{2-8} 
\multicolumn{1}{|c|}{}                             & \multicolumn{1}{c|}{T4}                & \multicolumn{1}{l|}{0.149  $\pm$  0.017} & \multicolumn{1}{l|}{0.204  $\pm$  0.022} & \multicolumn{1}{l|}{0.039  $\pm$  0.007} & \multicolumn{1}{l|}{0.650  $\pm$  0.007} & \multicolumn{1}{l|}{0.261  $\pm$  0.024} & \multicolumn{1}{l|}{0.884  $\pm$  0.008} \\ \cline{2-8} 
\multicolumn{1}{|c|}{}                             & \multicolumn{1}{c|}{T5}                & \multicolumn{1}{l|}{0.188  $\pm$  0.019} & \multicolumn{1}{l|}{0.242  $\pm$  0.023} & \multicolumn{1}{l|}{0.061  $\pm$  0.009} & \multicolumn{1}{l|}{0.662  $\pm$  0.007} & \multicolumn{1}{l|}{0.318  $\pm$  0.026} & \multicolumn{1}{l|}{0.893  $\pm$  0.008} \\ \cline{2-8} 
\multicolumn{1}{|c|}{}                             & \multicolumn{1}{c|}{T6}                & \multicolumn{1}{l|}{0.174  $\pm$  0.018} & \multicolumn{1}{l|}{0.234  $\pm$  0.023} & \multicolumn{1}{l|}{0.052  $\pm$  0.008} & \multicolumn{1}{l|}{0.652  $\pm$  0.007} & \multicolumn{1}{l|}{0.302  $\pm$  0.025} & \multicolumn{1}{l|}{0.903  $\pm$  0.008} \\ \cline{2-8} 
\multicolumn{1}{|c|}{}                             & \multicolumn{1}{c|}{T7}                & \multicolumn{1}{l|}{0.153  $\pm$  0.018} & \multicolumn{1}{l|}{0.207  $\pm$  0.022} & \multicolumn{1}{l|}{0.045  $\pm$  0.008} & \multicolumn{1}{l|}{0.644  $\pm$  0.007} & \multicolumn{1}{l|}{0.273  $\pm$  0.025} & \multicolumn{1}{l|}{0.880  $\pm$  0.008} \\ \cline{2-8} 
\multicolumn{1}{|c|}{}                             & \multicolumn{1}{c|}{T8}                & \multicolumn{1}{l|}{0.163  $\pm$  0.017} & \multicolumn{1}{l|}{0.218  $\pm$  0.022} & \multicolumn{1}{l|}{0.053  $\pm$  0.008} & \multicolumn{1}{l|}{0.658  $\pm$  0.007} & \multicolumn{1}{l|}{0.287  $\pm$  0.025} & \multicolumn{1}{l|}{0.885  $\pm$  0.008} \\ \hline
                                                   &                                        &                                          &                                          &                                          &                                          &                                          &                                          \\ \hline
\multicolumn{1}{|c|}{\multirow{8}{*}{InstructGPT}} & \multicolumn{1}{c|}{T1}                & \multicolumn{1}{l|}{0.114  $\pm$  0.015} & \multicolumn{1}{l|}{0.171  $\pm$  0.021} & \multicolumn{1}{l|}{0.029  $\pm$  0.006} & \multicolumn{1}{l|}{0.658  $\pm$  0.006} & \multicolumn{1}{l|}{0.229  $\pm$  0.024} & \multicolumn{1}{l|}{0.879  $\pm$  0.008} \\ \cline{2-8} 
\multicolumn{1}{|c|}{}                             & \multicolumn{1}{c|}{T2}                & \multicolumn{1}{l|}{0.125  $\pm$  0.017} & \multicolumn{1}{l|}{0.179  $\pm$  0.022} & \multicolumn{1}{l|}{0.037  $\pm$  0.007} & \multicolumn{1}{l|}{0.664  $\pm$  0.006} & \multicolumn{1}{l|}{0.235  $\pm$  0.024} & \multicolumn{1}{l|}{0.889  $\pm$  0.008} \\ \cline{2-8} 
\multicolumn{1}{|c|}{}                             & \multicolumn{1}{c|}{T3}                & \multicolumn{1}{l|}{0.119  $\pm$  0.016} & \multicolumn{1}{l|}{0.173  $\pm$  0.021} & \multicolumn{1}{l|}{0.034  $\pm$  0.007} & \multicolumn{1}{l|}{0.658  $\pm$  0.007} & \multicolumn{1}{l|}{0.229  $\pm$  0.025} & \multicolumn{1}{l|}{0.882  $\pm$  0.009} \\ \cline{2-8} 
\multicolumn{1}{|c|}{}                             & \multicolumn{1}{c|}{T4}                & \multicolumn{1}{l|}{0.109  $\pm$  0.015} & \multicolumn{1}{l|}{0.163  $\pm$  0.021} & \multicolumn{1}{l|}{0.028  $\pm$  0.006} & \multicolumn{1}{l|}{0.659  $\pm$  0.006} & \multicolumn{1}{l|}{0.225  $\pm$  0.024} & \multicolumn{1}{l|}{0.882  $\pm$  0.008} \\ \cline{2-8} 
\multicolumn{1}{|c|}{}                             & \multicolumn{1}{c|}{T5}                & \multicolumn{1}{l|}{0.123  $\pm$  0.016} & \multicolumn{1}{l|}{0.184  $\pm$  0.022} & \multicolumn{1}{l|}{0.033  $\pm$  0.007} & \multicolumn{1}{l|}{0.652  $\pm$  0.006} & \multicolumn{1}{l|}{0.241  $\pm$  0.024} & \multicolumn{1}{l|}{0.887  $\pm$  0.008} \\ \cline{2-8} 
\multicolumn{1}{|c|}{}                             & \multicolumn{1}{c|}{T6}                & \multicolumn{1}{l|}{0.122  $\pm$  0.016} & \multicolumn{1}{l|}{0.180  $\pm$  0.021} & \multicolumn{1}{l|}{0.033  $\pm$  0.006} & \multicolumn{1}{l|}{0.652  $\pm$  0.006} & \multicolumn{1}{l|}{0.244  $\pm$  0.025} & \multicolumn{1}{l|}{0.892  $\pm$  0.008} \\ \cline{2-8} 
\multicolumn{1}{|c|}{}                             & \multicolumn{1}{c|}{T7}                & \multicolumn{1}{l|}{0.105  $\pm$  0.014} & \multicolumn{1}{l|}{0.159  $\pm$  0.020} & \multicolumn{1}{l|}{0.024  $\pm$  0.005} & \multicolumn{1}{l|}{0.653  $\pm$  0.006} & \multicolumn{1}{l|}{0.221  $\pm$  0.024} & \multicolumn{1}{l|}{0.870  $\pm$  0.009} \\ \cline{2-8} 
\multicolumn{1}{|c|}{}                             & \multicolumn{1}{c|}{T8}                & \multicolumn{1}{l|}{0.109  $\pm$  0.016} & \multicolumn{1}{l|}{0.163  $\pm$  0.021} & \multicolumn{1}{l|}{0.031  $\pm$  0.007} & \multicolumn{1}{l|}{0.655  $\pm$  0.006} & \multicolumn{1}{l|}{0.223  $\pm$  0.025} & \multicolumn{1}{l|}{0.884  $\pm$  0.008} \\ \hline
                                                   &                                        &                                          &                                          &                                          &                                          &                                          &                                          \\ \hline
\multicolumn{1}{|c|}{\multirow{8}{*}{Llama2-7B}}   & \multicolumn{1}{c|}{T1}                & \multicolumn{1}{l|}{0.117  $\pm$  0.015} & \multicolumn{1}{l|}{0.160  $\pm$  0.020} & \multicolumn{1}{l|}{0.030  $\pm$  0.006} & \multicolumn{1}{l|}{0.674  $\pm$  0.006} & \multicolumn{1}{l|}{0.215  $\pm$  0.023} & \multicolumn{1}{l|}{0.885  $\pm$  0.009} \\ \cline{2-8} 
\multicolumn{1}{|c|}{}                             & \multicolumn{1}{c|}{T2}                & \multicolumn{1}{l|}{0.100  $\pm$  0.013} & \multicolumn{1}{l|}{0.143  $\pm$  0.019} & \multicolumn{1}{l|}{0.025  $\pm$  0.006} & \multicolumn{1}{l|}{0.675  $\pm$  0.006} & \multicolumn{1}{l|}{0.200  $\pm$  0.022} & \multicolumn{1}{l|}{0.875  $\pm$  0.009} \\ \cline{2-8} 
\multicolumn{1}{|c|}{}                             & \multicolumn{1}{c|}{T3}                & \multicolumn{1}{l|}{0.084  $\pm$  0.012} & \multicolumn{1}{l|}{0.113  $\pm$  0.015} & \multicolumn{1}{l|}{0.021  $\pm$  0.005} & \multicolumn{1}{l|}{0.681  $\pm$  0.006} & \multicolumn{1}{l|}{0.176  $\pm$  0.020} & \multicolumn{1}{l|}{0.879  $\pm$  0.009} \\ \cline{2-8} 
\multicolumn{1}{|c|}{}                             & \multicolumn{1}{c|}{T4}                & \multicolumn{1}{l|}{0.118  $\pm$  0.015} & \multicolumn{1}{l|}{0.156  $\pm$  0.019} & \multicolumn{1}{l|}{0.026  $\pm$  0.006} & \multicolumn{1}{l|}{0.678  $\pm$  0.006} & \multicolumn{1}{l|}{0.215  $\pm$  0.022} & \multicolumn{1}{l|}{0.892  $\pm$  0.008} \\ \cline{2-8} 
\multicolumn{1}{|c|}{}                             & \multicolumn{1}{c|}{T5}                & \multicolumn{1}{l|}{0.111  $\pm$  0.014} & \multicolumn{1}{l|}{0.142  $\pm$  0.017} & \multicolumn{1}{l|}{0.035  $\pm$  0.006} & \multicolumn{1}{l|}{0.673  $\pm$  0.007} & \multicolumn{1}{l|}{0.210  $\pm$  0.022} & \multicolumn{1}{l|}{0.876  $\pm$  0.009} \\ \cline{2-8} 
\multicolumn{1}{|c|}{}                             & \multicolumn{1}{c|}{T6}                & \multicolumn{1}{l|}{0.112  $\pm$  0.014} & \multicolumn{1}{l|}{0.146  $\pm$  0.017} & \multicolumn{1}{l|}{0.033  $\pm$  0.006} & \multicolumn{1}{l|}{0.674  $\pm$  0.007} & \multicolumn{1}{l|}{0.209  $\pm$  0.021} & \multicolumn{1}{l|}{0.872  $\pm$  0.009} \\ \cline{2-8} 
\multicolumn{1}{|c|}{}                             & \multicolumn{1}{c|}{T7}                & \multicolumn{1}{l|}{0.094  $\pm$  0.013} & \multicolumn{1}{l|}{0.122  $\pm$  0.016} & \multicolumn{1}{l|}{0.031  $\pm$  0.007} & \multicolumn{1}{l|}{0.693  $\pm$  0.006} & \multicolumn{1}{l|}{0.200  $\pm$  0.022} & \multicolumn{1}{l|}{0.880  $\pm$  0.009} \\ \cline{2-8} 
\multicolumn{1}{|c|}{}                             & \multicolumn{1}{c|}{T8}                & \multicolumn{1}{l|}{0.070  $\pm$  0.009} & \multicolumn{1}{l|}{0.094  $\pm$  0.012} & \multicolumn{1}{l|}{0.022  $\pm$  0.005} & \multicolumn{1}{l|}{0.699  $\pm$  0.006} & \multicolumn{1}{l|}{0.173  $\pm$  0.020} & \multicolumn{1}{l|}{0.888  $\pm$  0.009} \\ \hline
                                                   &                                        &                                          &                                          &                                          &                                          &                                          &                                          \\ \hline
\multicolumn{1}{|c|}{\multirow{8}{*}{Llama2-13B}}  & \multicolumn{1}{c|}{T1}                & \multicolumn{1}{l|}{0.110  $\pm$  0.015} & \multicolumn{1}{l|}{0.159  $\pm$  0.020} & \multicolumn{1}{l|}{0.027  $\pm$  0.006} & \multicolumn{1}{l|}{0.670  $\pm$  0.006} & \multicolumn{1}{l|}{0.222  $\pm$  0.023} & \multicolumn{1}{l|}{0.890  $\pm$  0.009} \\ \cline{2-8} 
\multicolumn{1}{|c|}{}                             & \multicolumn{1}{c|}{T2}                & \multicolumn{1}{l|}{0.104  $\pm$  0.013} & \multicolumn{1}{l|}{0.154  $\pm$  0.015} & \multicolumn{1}{l|}{0.031  $\pm$  0.006} & \multicolumn{1}{l|}{0.666  $\pm$  0.007} & \multicolumn{1}{l|}{0.222  $\pm$  0.022} & \multicolumn{1}{l|}{0.887  $\pm$  0.009} \\ \cline{2-8} 
\multicolumn{1}{|c|}{}                             & \multicolumn{1}{c|}{T3}                & \multicolumn{1}{l|}{0.106  $\pm$  0.012} & \multicolumn{1}{l|}{0.160  $\pm$  0.015} & \multicolumn{1}{l|}{0.026  $\pm$  0.005} & \multicolumn{1}{l|}{0.666  $\pm$  0.006} & \multicolumn{1}{l|}{0.228  $\pm$  0.020} & \multicolumn{1}{l|}{0.886  $\pm$  0.009} \\ \cline{2-8} 
\multicolumn{1}{|c|}{}                             & \multicolumn{1}{c|}{T4}                & \multicolumn{1}{l|}{0.113  $\pm$  0.015} & \multicolumn{1}{l|}{0.169  $\pm$  0.019} & \multicolumn{1}{l|}{0.028  $\pm$  0.006} & \multicolumn{1}{l|}{0.667  $\pm$  0.006} & \multicolumn{1}{l|}{0.234  $\pm$  0.021} & \multicolumn{1}{l|}{0.891  $\pm$  0.008} \\ \cline{2-8} 
\multicolumn{1}{|c|}{}                             & \multicolumn{1}{c|}{T5}                & \multicolumn{1}{l|}{0.084  $\pm$  0.014} & \multicolumn{1}{l|}{0.123  $\pm$  0.017} & \multicolumn{1}{l|}{0.024  $\pm$  0.005} & \multicolumn{1}{l|}{0.665  $\pm$  0.006} & \multicolumn{1}{l|}{0.190  $\pm$  0.022} & \multicolumn{1}{l|}{0.873  $\pm$  0.008} \\ \cline{2-8} 
\multicolumn{1}{|c|}{}                             & \multicolumn{1}{c|}{T6}                & \multicolumn{1}{l|}{0.080  $\pm$  0.014} & \multicolumn{1}{l|}{0.120  $\pm$  0.019} & \multicolumn{1}{l|}{0.021  $\pm$  0.006} & \multicolumn{1}{l|}{0.668  $\pm$  0.007} & \multicolumn{1}{l|}{0.180  $\pm$  0.022} & \multicolumn{1}{l|}{0.880  $\pm$  0.008} \\ \cline{2-8} 
\multicolumn{1}{|c|}{}                             & \multicolumn{1}{c|}{T7}                & \multicolumn{1}{l|}{0.091  $\pm$  0.012} & \multicolumn{1}{l|}{0.116  $\pm$  0.015} & \multicolumn{1}{l|}{0.026  $\pm$  0.006} & \multicolumn{1}{l|}{0.687  $\pm$  0.006} & \multicolumn{1}{l|}{0.180  $\pm$  0.020} & \multicolumn{1}{l|}{0.883  $\pm$  0.009} \\ \cline{2-8} 
\multicolumn{1}{|c|}{}                             & \multicolumn{1}{c|}{T8}                & \multicolumn{1}{l|}{0.082  $\pm$  0.011} & \multicolumn{1}{l|}{0.116  $\pm$  0.015} & \multicolumn{1}{l|}{0.026  $\pm$  0.006} & \multicolumn{1}{l|}{0.696  $\pm$  0.006} & \multicolumn{1}{l|}{0.196  $\pm$  0.021} & \multicolumn{1}{l|}{0.880  $\pm$  0.009} \\ \hline
\end{tabular}
\label{table:goodreads_full_results}
\end{table}

\end{document}